\documentclass[a4paper,12pt]{article}

\usepackage{amssymb,latexsym}

\usepackage{fullpage}

\makeatletter \@addtoreset{equation}{section} 
\makeatother

\newif\ifpdf \ifx\pdfoutput\undefined \pdffalse 

\else \pdfoutput=1 

\pdftrue \fi

\usepackage{graphicx} 

\begin{document} 
\begin{titlepage}
	\thispagestyle{empty} 
	
	\begin{flushright}
		
		\hfill{DFPD-07TH18}\\
		\hfill{CERN-PH-TH/2007-248} \\
		\hfill{ENSL-00196727}
	\end{flushright}
	
	\vspace{15pt} 
	\begin{center}
		{ \LARGE{\bf Gauged Supergravities \\[2mm]
		from \\[2mm]
		Twisted Doubled Tori \\[2mm]
		and \\[4mm]
		Non-Geometric String Backgrounds }}
		
		\vspace{30pt}
		
		\bf{Gianguido Dall'Agata$^\dagger$, Nikolaos Prezas$^\flat$, \\[2mm]
		Henning Samtleben$^\star$ and Mario Trigiante$^\natural$}
		
		\vspace{20pt}
		
		{\it $\dagger$ Dipartimento di Fisica ``Galileo Galilei'' $\&$ INFN, Sezione di Padova, \\
		Universit\`a di Padova, Via Marzolo 8, 35131 Padova, Italy}
		
		\vspace{20pt}
		
		{\it $\flat$ Physics Department, \\
		Theory Unit, CERN, \\
		CH-1211, Geneva 23, Switzerland }
		
		\vspace{20pt}
		
		{\it $\star$ Laboratoire de Physique, ENS-Lyon \\
		46, all\'ee d'Italie, F-69364 Lyon CEDEX07, France}
		
		\vspace{20pt}
		
		{\it $\natural$ Dipartimento di Fisica $\&$ INFN, Sezione di Torino, \\
		Politecnico di Torino , C.~so Duca degli Abruzzi, 24, I-10129 Torino, Italy }
		
		\vspace{20pt}
		
		{ABSTRACT} 
	\end{center}
	
	\vspace{10pt} We propose a universal geometric formulation of gauged supergravity in terms of a twisted doubled torus. 
	We focus on string theory (M-theory) reductions with generalized Scherk--Schwarz twists residing in the O$(n,n)$ (E$_{7(7)}$) duality group. 
	The set of doubled geometric fluxes, associated with the duality twists and identified naturally with the embedding tensor of gauged supergravity, captures all known fluxes, i.e.~physical form fluxes, ordinary geometric fluxes, as well as their non-geometric counterparts. 
	Furthermore, we propose a prescription for obtaining the effective geometry embedded in the string theory twisted doubled torus or in the M-theory megatorus and apply it for several models of geometric and non-geometric flux compactifications. 
\end{titlepage}
\baselineskip 6 mm

\section{Introduction} \label{sec:introduction}

Supergravity theories are usually studied as low-energy effective models of String Theory. 
Up to now, however, there are by far more possible supergravity realizations than those explained by String Theory reductions. 
The higher-dimensional origin of 4-dimensional supergravity models is especially important if we want to make contact between String Theory and phenomenology. 
Understanding which theories can be embedded into String Theory and which cannot, allows us to select consistent approaches for model building. 
Massive deformations of supergravity theories have a prominent position in this analysis. 
These theories allow for a stabilization of the moduli fields, which typically plague string effective theories, and for a non-trivial cosmological constant.

There are several ways of obtaining these supergravities from String Theory reductions. 
Compactifications on group manifolds, or certain coset spaces, give effective 4-dimensional actions that not only reproduce the original vacuum around which the theory was expanded, but also take into account certain deformations of the geometry of the internal space in a potential for the scalar fields. 
In these models, the gauge group of the effective theory is related to the symmetry group of the internal manifold.

Other approaches leading to gauged supergravities are Scherk--Schwarz reductions \cite{Scherk:1979zr}, which recently have received a new interpretation as compactifications on twisted tori \cite{Kaloper:1999yr}, and flux compactifications \cite{fluxesreview}. 
In these reductions the non-trivial structure of the effective theory gauge group follows from the couplings of the vector fields to the expectation values of the higher-dimensional form fields or to the torsion of the twisted manifold. 
Despite the great activity in the field, which consists in identifying and classifying effective supergravities arising from String Theory reductions that use the above mentioned approaches, it is clear that these cover only a percentage of the possible 4-dimensional realisations.

As often in the past, the use of duality relations on these effective theories and on the structure of the scalar potentials has revealed a structure larger than expected and the possibility of deriving more and new supergravity models by employing additional ``non-geometric'' fluxes \cite{Wecht:2007wu}. 
Although the existence of these fluxes can be argued from the effective theory data, we still cannot clearly explain their precise higher-dimensional origin. 
As the name suggests, these fluxes are often associated with reductions of String Theory on spaces that are not expected to have a global (or even local) geometric description. 
For this reason the task of verifying the validity regime of the resulting supergravities has become extremely challenging. 
In some cases we don't even know if there is a good way to describe String Theory on such backgrounds.

A step forward in trying to understand the origin of these models is given by the doubled formalism \cite{Duff:1989tf}--\cite{Hull:2007zu},
where, by doubling the coordinates of the internal space, one can obtain a description of String Theory and its effective models that allow for an explicit use of T-duality transformations. 
The basic idea underlying the doubled formalism is that, for genuinely stringy backgrounds such as the ``non-geometric'' ones, the winding modes of the string play an equally important role as the momentum modes and one should double the number of coordinates in order to account properly for both.

In this paper we aim to fill the gap existing between gauged supergravity theories and String Theory by providing a possible scheme for embedding any gauged supergravity theory with arbitrary gauge group (compatible, of course, with its embedding in the duality group) into a twisted version of the doubled formalism. 
This is a generalization of the Scherk--Schwarz reduction, known also as twisted torus compactification, to the case of doubled tori. 
The introduction of twist deformations to the ordinary reduction on the doubled torus gives a realization of all possible geometric and non-geometric fluxes, according to the type of coordinates and generators used in the twist matrices. 
The interesting feature of this approach is that one has information 
not only on the effective theory but also on the compactifying space. 
We discuss how the ordinary geometric data can be recovered from the twisted doubled geometry, by projections, and how the monodromies of the full space affect the global description of the reduced space.

In this paper we mainly focus on the ${\cal N} = 4$ models that can be derived by reduction of heterotic theory on a twisted doubled ${\mathbb T}^6$, models that give rise to effective theories with gauge groups whose representation on the gauge field strengths is embedded in the $O(6,6)$ T-duality group. 
However, a similar approach may be extended to more general models that lead to larger gauge groups embedded in the full duality group. 
For this reason we also discuss the extension of this approach to M-theory, proposing an embedding of all possible ${\cal N} = 8$ models into a ``twisted megatorus'' of dimension 56.

\bigskip

{\bf Note added:} While this paper was in preparation we received the preprint \cite{Hull:2007jy} whose discussion overlaps part of section \ref{sec:scherk_schwarz_on_a_doubled_torus}.

\section{Scherk--Schwarz on a doubled torus} 

\label{sec:scherk_schwarz_on_a_doubled_torus}

\subsection{Compactifications on twisted tori} 

\label{sub:subsection_name}

In \cite{Scherk:1979zr}, Scherk and Schwarz proposed a reduction scheme to obtain massive deformations of ordinary Kaluza--Klein reductions of gravity theories on tori by allowing a special non-trivial dependence of the spacetime fields on the internal coordinates. 
Since this dependence has eventually to disappear from the effective theory, it must be related to some symmetry acting on the lower-dimensional fields.

When reducing a theory of gravity on a $n$-dimensional torus ${\mathbb T}^n$, there is a natural SL$(n, {\mathbb R})$ symmetry acting on the fields of the effective action that can be used for such a purpose. 
Ordinary Kaluza--Klein expansion around a $\mathbb{T}^n$ compactification selects as 4-dimensional moduli the fluctuations $g_{ij}(x)$ around the flat torus metric\footnote{Zero fluctuation of the metric field means $g_{ij}(x) = \delta_{ij}$.}: 
\begin{equation}
	ds^{2} = g_{\mu\nu}(x) dx^{\mu} \otimes dx^{\nu} + g_{ij}(x) \left(dy^{i} + A_{\mu}^{i}dx^\mu\right) \otimes \left(dy^{j} + A_{\nu}^{j}dx^{\nu}\right). 
	\label{metrKK} 
\end{equation}
A Scherk--Schwarz reduction selects a different set of moduli, instead. 
Introducing a dependence of the fluctuations on the internal coordinates by a twist matrix $U^i{}_j(y)$, one obtains a new field expansion. 
For instance, for the metric field we have $g_{kl}(x,y)=g_{ij}'(x)U^i{}_k(y) U^j{}_l(y)$. 
The reduction Ansatz is then realized by an expansion in fluctuations around a non-trivial metric described by new vielbeins $\eta^i=U^i{}_j(y) dy^j$: 
\begin{equation}
	ds^{2} = g_{\mu\nu}(x) dx^{\mu} \otimes dx^{\nu} + g_{ij}'(x) \left(\eta^{i} + A_{\mu}^{i}dx^{\mu}\right) \otimes \left(\eta^{j} + A_{\nu}^{j}dx^{\nu}\right). 
	\label{metrSS} 
\end{equation}
These vielbeins describe a space that is a deformation of the original torus and for this reason such a reduction is also known as ``compactification on a twisted torus'' \cite{Kaloper:1999yr, Hull:2005hk}. 
The word ``twisted'' refers to the twisting of the frames $\eta^i$ with respect to the usual coordinate differentials $dy^i$ by the matrix $U^i{}_j(y) $. 
An equivalent way of thinking about this reduction is as a reduction on an ordinary torus but in the presence of a non-trivial spin connection condensate or torsion
\cite{Kaloper:1999yr, Hull:2005hk, Andrianopoli:2005jv}, therefore justifying the term ``geometric fluxes''.

The requirement that the $y$-dependance of the full Lagrangian is trivial upon inserting the reduction Ansatz (\ref{metrSS}) implies an important consistency condition: the vielbeins $\eta^i$ describe a group manifold, i.e. 
\begin{equation}
	d \eta^i = -\frac12 \, \tau^i_{jk}\, \eta^j \wedge \eta^k, \label{groupman} 
\end{equation}
for constant $\tau^i_{jk}$. 
As the reduction Ansatz (\ref{metrSS}) suggests, half of the isometry group of this group manifold, in this case the group of right-translations\footnote{Recall that the left-invariant vielbeins on a group manifold are dual to the left-invariant Killing vectors that generate right translations.}, becomes the gauge group of the effective lower-dimensional theory. 
Consequently the $\tau$-constants appearing in (\ref{groupman}) are the structure constants of the gauge group 
\begin{equation}
	[Z_i, Z_j] = \tau^k_{ij} Z_k . 
	\label{gaugegroup} 
\end{equation}
Moreover, twisting the torus induces a potential for the Scherk--Schwarz moduli fields $g'_{ij}(x)$ given by \cite{Scherk:1979zr} 
\begin{equation}
	V = 2\, \tau^i_{jk} \,\tau^j_{il}\, g'^{kl} + \tau^i_{jk} \,\tau^m_{np}\, g'_{im} g'^{jn} g'^{kp}. 
	\label{sspot} 
\end{equation}

Formula (\ref{gaugegroup}) implies that there is a constructive way to define the twist matrices and hence the reduction Ansatz for a given gauge group ${\cal G} \subset {\rm SL}(n,{\mathbb R})$. 
One starts by selecting a group representative $g(y) = {\rm exp}(y^i Z_i) \in {\cal G}$, where $Z_i$ are the generators of the algebra ${\mathfrak g}$ (\ref{gaugegroup}). 
Then one extracts the vielbeins $\eta^i$ by inspection of the left-invariant Maurer--Cartan form $\Omega = g^{-1}dg = \eta^i Z_i$. 
Finally, in order for the theory to be consistently defined on a compact space, one considers only groups such that a left quotient $\Gamma\backslash{\cal G}$ with respect to the compact subgroup $\Gamma = {\cal G}({\mathbb Z})$ is possible.

When reducing string theory in such a way, one has to take into account other spacetime fields besides the metric \cite{Scherk:1979zr,Kaloper:1999yr,Hull:2005hk}. 
The same type of reduction can be extended to all other string theory fields, like the universally present 2-form field $B$ or other higher rank form fields. 
The resulting supergravity theory is still going to be a gauged supergravity, its gauge group, however, is not going to be just ${\cal G}$ due to the existence of extra vector fields coming from the reduction of the form fields (for instance $B_{\mu i}$ from the universal 2-form). 
Typically, the twisting induces a non-abelian action of the gauge symmetries also on them and therefore the final gauge group will be generically different than ${\cal G}$ \cite{Scherk:1979zr,Kaloper:1999yr, Hull:2005hk},\cite{Angelantonj:2003rq}--\cite{D'Auria:2005rv}.

The previous remark explains why a collective description of the lower dimensional moduli in a single generalized metric is desirable. 
A construction similar to the above one for this generalized metric would immediately give rise to an effective supergravity theory with the right gauge group described by the group manifold on which one reduces the original theory. 
In the following section we make a proposal on how to obtain such a description from twisting the doubled torus of \cite{Hull:2004in}.

\subsection{Twisting the doubled torus} 

\label{sub:twisting_the_doubled_torus}

When reducing string theory to four dimensions one has several moduli fields coming from different sources: the metric as well as the various form fields appearing in the ten-dimensional theory. 
A collective description of them in a unique generalized metric would be desirable, but for simplicity one can focus first only on the common sector of all string theory models described by the metric and the $B$-field system. 
In the reduction of the common sector to 4 dimensions one finds that the moduli fields describe the following non-linear $\sigma$-model \cite{Maharana:1992my}: 
\begin{equation}
	\frac{{\rm SO}(6,6)}{{\rm SO}(6)\times {\rm SO}(6)}. 
	\label{smodel} 
\end{equation}

The 36 degrees of freedom parametrizing the above coset consist of the fluctuations of the metric $g_{ij}$ and of the 2-form field $B_{ij}$ in the internal space. 
These can be collected in a suggestive O$(6,6)$ matrix 
\begin{equation}
	{\cal H}_{IJ} = \left( 
	\begin{array}{cc}
		g_{ij} - B_{ik}g^{kl}B_{lm} & B_{ik}g^{kl}\\
		-g^{jk}B_{kl} & g^{ij} 
	\end{array}
	\right) . 
	\label{genmetr} 
\end{equation}
This matrix has the right properties to be interpreted as the generalized metric of a doubled internal space with coordinates $${\mathbb Y}^M = \{y^i ,\widetilde{y}_i\},$$ so that SO$(6,6)$ has a natural action on them \cite{Duff:1989tf}. 
In the case of torus compactifications, a generalized world-sheet action has been proposed for strings living in this doubled space \cite{Hull:2004in,Hull:2006va}.

In this doubled formalism, the SO$(6,6)$ group has a clear interpretation as T-duality transformations on the background and the doubled coordinates have also an obvious interpretation as coordinates on the original and the T-dual circles of the internal ${\mathbb T}^6$. 
Since ${\cal H}$ defined in (\ref{genmetr}) appears in this formalism as a metric for the doubled torus it is somehow natural to interpret the effective 4-dimensional theories coming from this type of compactifications as reductions on a doubled internal space: 
\begin{equation}
	d{\cal S}^2 = {\cal H}_{MN}\big(B(x),g(x)\big) \left(d{\mathbb Y}^M + A_\mu^M dx^\mu\right)\otimes\left(d{\mathbb Y}^N + A_\nu^N dx^\nu \right). 
	\label{doubledmetr} 
\end{equation}
The vector fields $A_\mu^M$ comprise those coming from the metric and the $B$-field: $A_\mu^M = \{g_\mu^i, B_{\mu i}\}$ and there is a natural action of the duality group on them. 
The resulting effective theory is an ordinary (i.e.~non-gauged) supergravity with gauge group ${\cal G}=$ U$(1)^{12}$.

Now, if we draw a parallel to the previous discussion where a Scherk--Schwarz reduction led to a non-abelian gauge group ${\cal G}$ by introducing a twist matrix between the metric moduli and the differential on the internal space, we are tempted to generalize (\ref{doubledmetr}) by introducing a \emph{twisted doubled torus}: 
\begin{equation}
	d{\cal S}^2 = {\cal H}_{MN}\left(B(x),g(x)\right) \left(U^M{}_P({\mathbb Y})d{\mathbb Y}^P + A_\mu^M dx^\mu\right)\otimes\left(U^N{}_Q({\mathbb Y})d{\mathbb Y}^Q + A_\nu^N dx^\nu \right). 
	\label{twistedSS} 
\end{equation}
In general, these twist matrices $U$ may depend on both the ordinary and the dual coordinates and soon we will see how this may lead to deformations of the original torus which may not have a global (or even a local) geometric description.

As before, we expect that this explicit dependence cancels in the effective theory provided these twists define proper generalized vielbeins on $T^*({\mathbb T}^{12})$ 
\begin{equation}
	{\mathbb E}^M = U^M{}_N({\mathbb Y})\, d{\mathbb Y}^N, \label{genvielb} 
\end{equation}
such that 
\begin{equation}
	d {\mathbb E}^M =-\frac12\, {\cal T}_{NP}{}^M \, {\mathbb E}^N \wedge {\mathbb E}^P \label{torsions} 
\end{equation}
for constant ${\cal T}$. 
These constants can be viewed as a sort of ``generalized geometric fluxes'' and it is natural to expect that they will include geometric fluxes, ordinary physical form-fluxes, but even the ``non-geometric'' fluxes proposed in \cite{ Shelton:2005cf, Aldazabal:2006up, Shelton:2006fd,Lawrence:2006ma, Wecht:2007wu}. 
Therefore, it should be identified with the embedding tensor of gauged supergravity. 
One can also motivate the twisting of the doubled torus by comparing the formula (\ref{sspot}) for the potential of a Scherk--Schwarz reduction on a standard twisted torus to the scalar potential of a generic gauged supergravity, which schematically reads 
\begin{equation}
	V \sim 2 \, {\cal T}_{JK}{}^I\, {\cal T}_{IL}{}^J\, {\cal H}^{KL} + {\cal T}_{JK}{}^I \,{\cal T}_{NP}{}^M \,{\cal H}_{IM} {\cal H}^{JN} {\cal H}^{KP}. 
	\label{ssgenpot} 
\end{equation}
We see that the generalized metric indeed plays the role of metric moduli for the twisted doubled torus and the embedding tensor corresponds to geometric flux for the doubled torus.

The 4-dimensional fields encoded in (\ref{genmetr}) can be interpreted as the fluctuations around a generalized background metric 
\begin{equation}
	{\mathfrak H}_{MN}({\mathbb Y}) = U^P{}_M({\mathbb Y}) \delta_{PQ} U^Q{}_N({\mathbb Y}), \label{generalized} 
\end{equation}
where $\delta_{KL} = {\rm diag}\{++\ldots+\}$ defines the SO(6)$\times$SO(6) invariant metric. 
As in the case of the ordinary Scherk--Schwarz reduction, the vielbeins (\ref{genvielb}) can be viewed as vielbeins on a group manifold ${\cal G}$, properly compactified by the action of a discrete group $\Gamma = {\cal G}(\mathbb Z)$. 
In this case, however, the group ${\cal G}$ \emph{is} the full gauge group of the effective theory, since the action on the vector fields contained in (\ref{twistedSS}) includes all the vectors of the effective theory.

At this point it is natural to ask ourselves what kind of groups can be used in this reduction. 
By analogy with the ordinary Scherk--Schwarz reduction one would expect ${\cal G} \subset {\rm GL}(12,{\mathbb R})$. 
On the other hand, since the actual duality group is SO(6,6) one would expect ${\cal G} \subset$ SO(6,6). 
The correct answer is in between. 
As we will explain in the next section, the faithful representation of the gauge group realized on the vector fields of any gauged supergravity may actually correspond to an algebra that is larger than the one realized on the curvatures. 
Therefore, the structure of the gauge group read from the commutators on the vector fields may be larger than the one that has to be embedded in the O(6,6) duality group, as would naturally follow by the Gaillard--Zumino prescription \cite{Gaillard:1981rj}. 
This interesting feature is actually true in full generality, even for ordinary Scherk--Schwarz reductions, and therefore we discuss it in the next section without requiring at first any link to the doubled formalism.

\subsection{Fluxes and symplectic embedding} 

\label{sub:fluxes_and_symplectic_embedding}

When performing Scherk--Schwarz compactifications there is often a mismatch between what is called the gauge group of the effective theory and the group embedded in the duality symmetries group. 
The gauge group obtained from Scherk--Schwarz reductions can be read from the actions on the vector fields, which should form a faithful representation of this group. 
If we call $\widehat{X}_M$ the generators of the corresponding algebra in the faithful representation and $X_M$ those in the adjoint, the latter being embeddable in the duality algebra, one sees that the group generated by $X_M$ is a contraction of that generated by $\widehat{X}_M$ . 
More precisely, the abelian ideal ${\mathfrak I}$, comprising of generators which act trivially on the curvatures, has been removed: 
\begin{equation}
	\widetilde{\cal G} = {\cal G}/{\mathfrak I}. 
	\label{contract} 
\end{equation}

To be more explicit, take the action of the gauge transformations on the vector fields and their covariant field strengths: 
\begin{eqnarray}
	\delta_\epsilon A^M & = & d \epsilon^M - {\cal T}_{NP}{}^M \epsilon^N A^P, \label{gaugegrpu} \\[2mm]
	\delta_\epsilon F^M & = & -{\cal T}_{NP}{}^M \epsilon^N F^P. 
	\label{gaugegrpuF} 
\end{eqnarray}
If one computes the commutator of two such transformations on the gauge vectors, forming a faithful representation of the gauge group, one gets 
\begin{equation}
	\left[\delta_{\epsilon_1},\delta_{\epsilon_2}\right] A^M = \delta_{[\epsilon_1,\epsilon_2]} A^M, \label{comm} 
\end{equation}
and therefore one always obtains a non-zero result for non-commuting generators. 
On the other hand, if one takes the same commutator on the field-strengths one gets a zero on the right hand side any time the commutator closes on a generator of $\widehat{X}$ which is not  $X$, i.e.~any generator which has a trivial action on the curvatures.

Let us illustrate the above point by considering the case of the Heisenberg group 
\begin{equation}
	\left[\widehat{X}_1, \widehat{X}_2 \right] = \widehat{X}_3. 
	\label{Heis} 
\end{equation}
This can be realized on the gauge field strengths as 
\begin{eqnarray}
	\delta A^1 &=& d \epsilon^1, \\[2mm]
	\delta A^2 &=& d \epsilon^2, \\[2mm]
	\delta A^3 &=& d \epsilon^3 + \epsilon^1 A^2 - \epsilon^2 A^1. 
\end{eqnarray}
It is trivial to show that the commutator (\ref{Heis}) on $A^3$ correctly gives the action of $\widehat{X}_3$ on the same vector. 
This action is faithful only because of the inhomogeneous term in the transformation rules of the connections. 
On the field strengths one gets 
\begin{eqnarray}
	\delta F^1 &=& 0, \\[2mm]
	\delta F^2 &=& 0, \\[2mm]
	\delta F^3 &=& \epsilon^1 F^2 - \epsilon^2 F^1. 
\end{eqnarray}
Despite the non-trivial action on $F^3$, the commutator of the $X_1$ and $X_2$ generators on $F^3$ closes to zero giving 
\begin{equation}
	\left[X_1,X_2\right] =0. 
	\label{comm2} 
\end{equation}
The action of $X_3$ is trivial: $X_3 = 0$. 
This is the adjoint action and for this action the generators commute and can be embedded in the duality group. 
This explains why often the full gauge group ${\cal G}$ cannot be embedded in the duality group. 
Only a contraction, where the generators of the ideal ${\mathfrak I}$ have been set to zero, can be effectively embedded in the duality group. 
The most extreme manifestation of this phenomenon is the case of an abelian gauge group U(1)$^{12}$; obviously the faithful representation of that group cannot be embedded in the fundamental of O(6,6) whereas the adjoint one, consisting of trivial generators equal to zero, admits also a trivial embedding. 
More interesting gauge algebras exhibiting this property were found in flux compactifications of String or M-theory \cite{Angelantonj:2003rq,Angelantonj:2003up,Dall'Agata:2005ff}.

Coming back to the doubled formalism, all of the above applies also to the twisted doubled torus which is a discrete quotient $\Gamma\backslash{\cal G}$ of a group manifold ${\cal G}$. 
The group $\mathcal{G}$ may be non-semisimple and thus its adjoint representation may not be faithful (for instance, as we saw above, if the corresponding Lie algebra $\frak{g}$ has central elements, the adjoint representation of these would be trivial). 
This affects the construction of the vielbein (\ref{genvielb}) on $\mathcal{G}$ as follows.

Let $\widehat{X}_M$ be generators of the Lie algebra $\frak{g}$ of $\mathcal{G}$ in a faithful representation, satisfying: 
\begin{eqnarray}
	[\widehat{X}_M,\,\widehat{X}_N]&=&{\cal T}_{MN}{}^P\,\widehat{X}_P\,. 
	\label{faithf} 
\end{eqnarray}
The $X_M$ denote a basis of generators of $\frak{g}$ in the adjoint representation: 
\begin{eqnarray}
	(X_M)_N{}^P&=&{\cal T}_{NM}{}^P\,. 
\end{eqnarray}
The group ${\cal G}$ is the gauge group of the effective gauged supergravity and therefore the adjoint representation of $\frak{g}$ has to be embedded in the fundamental representation of the duality algebra $\frak{o}(6,6)$. 
This means that the $X_M$ generators can be expressed as linear combinations of the generators $t_\alpha$ of the fundamental representation of the duality group
\begin{eqnarray}
	X_M&=&\theta_M{}^\alpha\,t_\alpha, \label{embedd} 
\end{eqnarray}
through the embedding tensor $\theta_M{}^\alpha$. 
This requirement implies that on the tangent space of ${\cal G}$ one can define an O$(6,6)$ invariant metric ${\cal I}$: 
\begin{equation}
	{\cal I} = \left( 
	\begin{array}{cc}
		0 & 1_6\\
		1_6 & 0 
	\end{array}
	\right) \label{invmetric} 
\end{equation}
The existence of this metric will be crucial when we will try to give an ordinary spacetime interpretation to the internal manifold.

It can now be seen how the doubled vielbein follows from a generic element of $\mathcal{G}$ that is constructed from the faithful representation (\ref{faithf}): 
\begin{eqnarray}
	g({\mathbb Y})&=& {\rm e}^{{\mathbb Y}^M\,\widehat{X}_M}\,. 
\end{eqnarray}
The left-invariant one-form determines the vielbein ${\mathbb E}^M$: 
\begin{eqnarray}
	g^{-1}dg&=&{\mathbb E}^M\,\widehat{X}_M=U^N{}_M\,\widehat{X}_N\,d{\mathbb Y}^M\,. 
\end{eqnarray}
Moreover, from the same construction one can also determine explicitly this vielbein and especially the ``twist matrix'' $U^N{}_M({\mathbb Y})$. 
It is very interesting to notice that it can be expressed in terms of the generators in the adjoint representation $X_M$ only: 
\begin{eqnarray}
	U^N{}_M&=&\left({\mathbf 1}+\sum_{k=1}^\infty\frac{1}{(k+1)!}\,({\mathbb Y}^P\,X_P)^k\right)^N{}_M\,.\label{nonexp} 
\end{eqnarray}

\subsection{Ordinary geometry from a doubled space} 

\label{sub:ordinary_geometry_from_a_doubled_space}

So far we mainly discussed how a twisted version of the doubled geometry may lead to general gauged supergravities that include all known models. 
In this section we want to address the problem of understanding the geometry described by the doubled vielbeins ${\mathbb E}^M$ in terms of an ordinary 6-dimensional space.

Before the twisting, the doubled space is that of a torus ${\mathbb T}^{12}$ that is a trivial fibration of one ${\mathbb T}^6$ over another ${\mathbb T}^6$. 
When the twist is introduced, the new doubled space described by the ${\cal H}_{MN}$ metric is a (compact) group manifold of dimension 12. 
However, this total space can still be described as a fibration of a 6-dimensional space on another, provided we manage to describe properly the split into a base and fiber. 
The existence of an O(6,6)-invariant metric ${\cal I}_{MN}$ (\ref{invmetric}) on the tangent space of ${\cal M}_{12}=\Gamma\backslash{\cal G}$ comes to the rescue\footnote{Notice that we are mostly interested in the case $d=6$ but the discussion below is general and does not depend on the actual value of $d$.}. 
As we saw in (\ref{embedd}), the adjoint representation of the algebra generating the group manifold ${\cal G}$ has to be embedded in the fundamental representation of the duality group. 
This means that the vielbeins, which transform in the co-adjoint of ${\cal G}$, transform also in the (dual of the) fundamental of O$(6,6)$. 
Then, the O$(6,6)$ invariant metric can be used to define an inner product on the cotangent space of ${\cal M}_{12}$ 
\begin{equation}
	\langle {\mathbb E}^M, {\mathbb E}^N\rangle = {\cal I}^{MN}, \label{intprod} 
\end{equation}
so that locally $T^*({\cal M}_{12})$ splits into the sum of a tangent and cotangent space on a 6-dimensional space $$T^*({\cal M}_{12}) = T^*({\cal M}_{6})+ T({\cal M}_{6}).$$ 
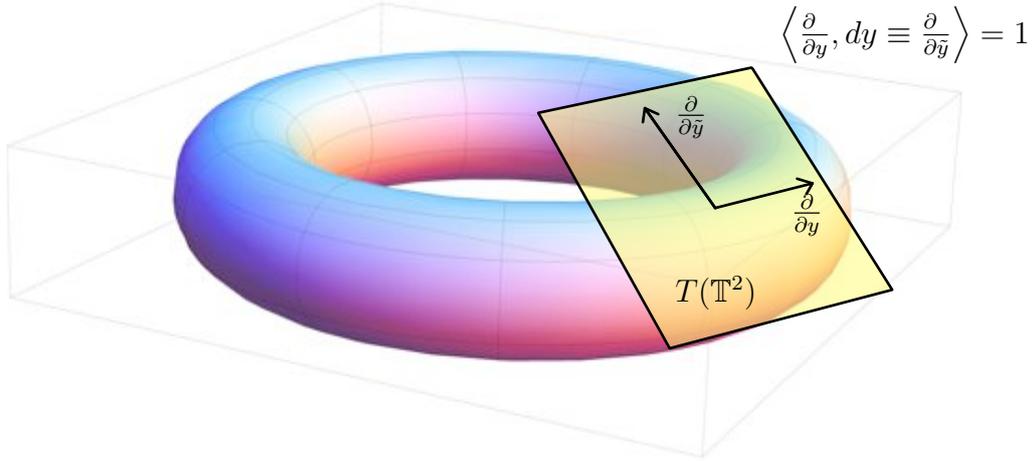
\begin{figure}[htb]
	\setlength{\unitlength}{3in} 
	\begin{center}
		\begin{picture}
			(1.,1.) \put(-.5,.0){ 
			\includegraphics[scale=1]{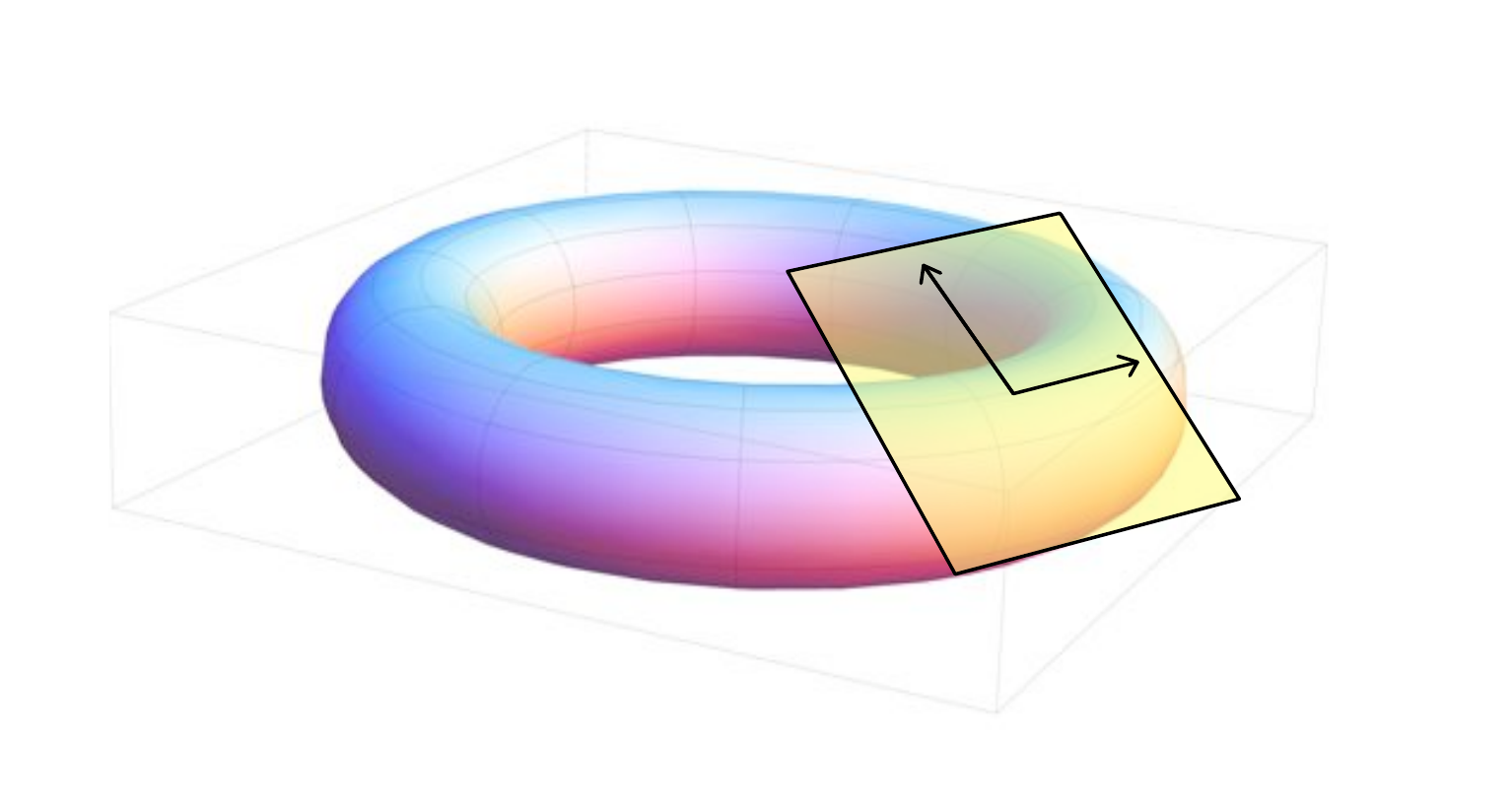} } \put(1.02,.54){$\frac{\partial}{\partial y}$} \put(.82,.71){$\frac{\partial}{\partial \tilde y}$} \put(1,.85){$ \left\langle\frac{\partial\phantom{y}}{\partial y},dy \equiv \frac{\partial\phantom{y}}{\partial \tilde y}\right\rangle = 1 $} \put(.82,.4){$T({\mathbb T}^2)$} 
		\end{picture}
		\caption{\it The tangent space on a torus $T({\mathbb T}^2) = \{\lambda \frac{\partial}{\partial y} + \mu \frac{\partial}{\partial \widetilde y}, \lambda,\mu \in {\mathbb R}\}$ splits into the tangent space on its base circle $T({\mathbb S}^1) = \{\lambda \frac{\partial}{\partial y}, \lambda \in {\mathbb R}\}$ plus its cotangent space $T^*({\mathbb S}^1) = \{\lambda dy \simeq \lambda \frac{\partial}{\partial \widetilde y}, \lambda \in {\mathbb R}\}$ when there is a natural O(1,1) inner product pairing $\frac{\partial}{\partial y}$ and $\frac{\partial}{\partial \widetilde y}$.} \label{fig:torodoub} 
	\end{center}
\end{figure}

The inner product (\ref{intprod}) yields an almost product structure that for twists in O(6,6) can be applied also to the basis of differentials $d{\mathbb Y}$. 
Using (\ref{genvielb}) we can see that the inner product inherited by the basis of differentials is 
\begin{equation}
	\langle d{\mathbb Y},d{\mathbb Y}\rangle = U^{-1} {\cal I} \left(U^{-1}\right)^T \equiv \gamma^{-1}({\mathbb Y}). 
	\label{inhint} 
\end{equation}
If the twists $U$ are in O(6,6) the metric $\gamma$ is actually identical to ${\cal I}$. 
This implies that one can really consider the differentials of the dual coordinates $d\widetilde y_i$ as the basis of vectors on the tangent space of the manifold ${\cal M}_6$ described by the $y^i$ coordinates (see Fig.~1) : 
\begin{equation}
	\langle dy^i, d\widetilde y_j\rangle = \delta^i_j \qquad \Rightarrow \qquad d\widetilde y_i = \frac{\partial\phantom{y}}{\partial y^i}. 
	\label{duali} 
\end{equation}
Moreover, if $U \in \, $O(6,6) the doubled vielbeins can be put in the triangular form 
\begin{equation}
	{\mathbb E} = \left( 
	\begin{array}{l}
		dy^i e_i^{a} \\[2mm]
		d\widetilde y_i (e^{-1})^i_a + dy^j B_{ji}(e^{-1})^i_a 
	\end{array}
	\right) \label{triangular} 
\end{equation}
by using an O(6)$\times$O(6) local transformation. 
The doubled space can then be interpreted as following from an ``ordinary'' background described by a metric and a B-field given by (\ref{triangular}). 
This does not yet imply that a well-defined global geometric picture exists, because, as we will see later, global issues may spoil this description. 
Anyway, it is important to understand for which group manifolds we have such a description and therefore which gauged supergravities can be lifted to string theory using this type of backgrounds.

Upon inspection of (\ref{nonexp}), one realizes that for a generic choice of ${\mathcal G}$, the matrix $U$ is not, in general, the exponential of generators of O$(6,6)$ and therefore it is not a group element of O$(6,6)$. 
However, there are instances where the right hand side of (\ref{nonexp}) becomes an O$(6,6)$ group element. 
For example, if the $X_M$ generators (in the adjoint representation) are nilpotent of order 2, namely $X_M\cdot X_N=0$, in other words the gauge algebra is solvable of order 2,
then $U$ can be trivially written as an ${\rm O}(6,6)$ group element 
\begin{eqnarray}
	U&=& {\mathbf 1} +\frac{1}{2}\,{\mathbb Y}^M X_M={\rm e}^{\frac{1}{2}\,{\mathbb Y}^M X_M}. 
\end{eqnarray}
There may be other non-trivial solutions but it is clearly difficult to provide examples just by analyzing (\ref{nonexp}). 
It is therefore better to have an intrinsic characterization of the situations in which $U \in $O(6,6), possibly after reparametrizations of ${\mathbb Y}$. 
This is provided by the following criterion: \emph{the twist matrix $U$ can be brought to an ${\rm O}(n,n)$ matrix through a reparametrization if and only if the Riemann tensor constructed out of the metric $\gamma_{MN}$ defined in (\ref{inhint}) vanishes: } 
\begin{equation}
	R_{MNP}{}^Q(\gamma) = 0. 
	\label{condRiem} 
\end{equation}

First of all, it should be clear that this Riemann tensor is not related to the curvature of the group manifold computed from the ${\cal H}$ metric, which follows from ${\mathbb E}^M \otimes {\mathbb E}^N \delta_{MN}$, instead. 
Then, whenever the Riemann tensor constructed from the metric $\gamma_{MN}$ vanishes, the space defined by this metric is flat and therefore there is a change of coordinates which brings $\gamma_{MN} = {\cal I}_{MN}$. 
The important ingredient is that for the doubled space described by the discrete quotient of the group manifold ${\cal G}$ this computation can be done without any explicit reference to the metric $\gamma$. 
Since a group manifold is homogeneous, its Riemann tensor is constant in rigid indices and can be expressed in terms of the structure constants of the group itself. 
If the structure constants are denoted by ${\cal T}_{MN}{}^P=-X_{MN}{}^P$, the doubled vielbeins satisfy 
\begin{eqnarray}
	d{\mathbb E}^P+\frac{1}{2}\,{\cal T}_{MN}{}^P{\mathbb E}^M\wedge {\mathbb E}^N&=&d{\mathbb E}^P+\omega_N{}^P\wedge {\mathbb E}^N=0\,. 
\end{eqnarray}
The spin connection then has the form 
\begin{eqnarray}
	\omega_{M,N}{}^P&=&\frac{1}{2}\,\left({\cal T}_{MN}{}^P+{\cal I}^{PP^\prime}\,{\cal I}_{MM^\prime}{\cal T}_{P^\prime N}{}^{M^\prime}+{\cal I}^{PP^\prime}\,{\cal I}_{NM^\prime}{\cal T}_{P^\prime M}{}^{M^\prime}\right) = \frac12 {\cal T}_{MN}{}^P\,, 
\end{eqnarray}
where the last equality follows from the adjoint representation being embedded in O(6,6), and the Riemann tensor reads: 
\begin{eqnarray}
	R_M{}^N&=&d\omega_M{}^N-\omega_M{}^P\wedge \omega_P{}^N=-\frac{1}{2}\,R_{PQ,M}{}^N\,{\mathbb E}^P\wedge {\mathbb E}^Q\,,\nonumber\\[2mm]
	R_{PQ,M}{}^N&=&\omega_{PM}{}^R\,\omega_{QR}{}^N-\omega_{QM}{}^R\,\omega_{PR}{}^N+{\cal T}_{PQ}{}^R\,\omega_{RM}{}^N \\[2mm]
	&=& \frac14 {\cal T}_{PQ}{}^R {\cal T}_{RM}{}^N. 
	\nonumber 
\end{eqnarray}
This allows a classification of the group manifolds that can be brought to the form of an ordinary doubled torus geometry with twist matrices that can be deduced from a metric and $B$-field, just by using the intrinsic characterization given by its structure constants. 
It is actually striking that only solvable algebras
of order two satisfy the above relation for any choice of the indices: ${\cal T}_{MN}{}^P {\cal T}_{PQ}{}^R =- (X_M X_Q)_N{}^R= 0$.

In the remainder of this section we come back to the examples where the product of the differentials (\ref{inhint}) is not constant and therefore cannot be equal to ${\cal I}$. 
When this happens, we can still try to make sense of the resulting space by obtaining a vielbein representation of the form (\ref{triangular}). 
When $\langle dy^i, d\widetilde y_j\rangle = \delta^i_j$, the identification of the differentials of the dual coordinates as the dual basis to $\{dy^i\}$ gives a proper coordinate basis on the base space for its tangent and cotangent space. 
If, on the other hand, the inner product $\langle d{\mathbb Y}^M,d{\mathbb Y}^N\rangle = \gamma^{MN}({\mathbb Y}) $ gives a coordinate dependent result, with a metric $\gamma$ that cannot be brought to a constant form by a coordinate transformation, the splitting of $T^*({\cal M}_{12})$ into $T^*({\cal M}_{6})+ T({\cal M}_{6})$ has to be done pointwise. 
This means that the definition of the vector field basis $\{\partial_i\}$ in terms of $\{d\widetilde y_i\}$ has to have a coordinate dependent form. 
If, for instance, the metric $\gamma$ is block-diagonal but non-trivial, i.e. 
\begin{equation}
	\gamma = \left( 
	\begin{array}{cc}
		0 & {\gamma^i}_j({\mathbb Y}) \\[2mm]
		{\gamma_i}^j({\mathbb Y}) & 0 
	\end{array}
	\right) , \label{nontrgam} 
\end{equation}
one can define 
\begin{equation}
	\partial_i \equiv (\gamma^{-1})_i{}^j d \widetilde y_j , 
\end{equation}
so that 
\begin{equation}
	\left\langle dy^i, \frac{\partial\phantom{y}}{\partial y^j}\right\rangle = \delta^i_j. 
	\label{properinn} 
\end{equation}
Defining ${\mathbb E}^M$ in terms of the $\{dy^i, \partial_i\}$ basis gives an O(6,6) matrix which, therefore, can be put in a triangular gauge as 
\begin{equation}
	{\mathbb E} = \left( 
	\begin{array}{l}
		dy^i e_i^{a} \\[2mm]
		\displaystyle (e^{-1})^i_a \frac{\partial\phantom{y}}{\partial y^i}+ dy^j B_{ji}(e^{-1})^i_a 
	\end{array}
	\right). 
	\label{triangular2} 
\end{equation}
We will  discuss later one such example where we will double the twisted torus corresponding to the flat group gauging on an ordinary ${\mathbb T}^6$.

In order to complete the interpretation of the base space geometry, besides global issues which will be discussed in the next section, one has still to make sense of the dependence on the dual coordinates. 
The explicit presence of dual coordinates $\widetilde y_i$ in the resulting metric and $B$-field may be a signal that we are actually dealing with a non-geometric space. 
From the point of view of the base space we expect these dual coordinates to be interpreted as non-local loop coordinates. 
However, it may happen that a gauge transformation allows us to get rid of them when going to the triangular gauge. 
Also, as we will see in the next section, we can consider setting them to zero, provided we take into account the effect of the monodromies involving them as an action on the fields of the effective theory. 
After all, the ordinary space is obtained by a projection from the global doubled torus to the 6-dimensional base parametrized by the $y^i$ coordinates and whose cotangent space is spanned by the $E^i$ vielbeins. 
Hence, the projection is also performed by setting $\widetilde y_i$ to zero. 
We refer the reader to the next section for a concrete description of this mechanism, while we hope to return on this issue with more details elsewhere.

\subsection{Global description} 

\label{sub:global_description}

In the previous subsections we described how the doubled geometry can be interpreted as an ordinary 6-dimensional background with a metric and a $B$-field, provided certain conditions are met. 
In this section we want to discuss the effects of global constraints on the simple description of the doubled space given in the previous subsections.

First of all, in order to have a globally well-defined vielbein on the full compact space $\Gamma\backslash{\cal G}$, one needs to identify the coordinates as 
\begin{equation}
	{\mathbb Y} \sim \sigma_L (g,{\mathbb Y}), \label{identif} 
\end{equation}
where $\sigma_L (g,{\mathbb Y})$ is the left action of an arbitrary group element $g \in \Gamma$ (generated by right invariant vector fields). 
Since ${\mathbb E}$ are left-invariant 
\begin{equation}
	{\mathbb E}(\sigma_L(g,{\mathbb Y}))=\,{\mathbb E}({\mathbb Y}). 
	\label{compa} 
\end{equation}
In full generality, such an identification will mix the coordinates $y^i$ with the dual ones $\widetilde y_i$. 
Clearly, when trying to obtain a description in terms of an ordinary twisted torus metric with a non-trivial $B$-field the dual coordinates should disappear and one should set them to zero in the process of projecting to the base space. 
On the other hand, this procedure may not be strictly compatible with the identifications obtained in (\ref{identif}). 
However we can still obtain a consistent picture by considering the action of the monodromy group for fixed dual coordinates and then take into account its effect on the base space by translating the action on the coordinates into an action on the fields of the effective theory.

The generalized vielbein ${\mathbb E}$ transforms in the co-adjoint representation of $\mathcal{G}$ and this is what defines the \emph{monodromy} of the compact manifold $\Gamma\backslash\mathcal{G}$. 
If we denote by $\sigma_R (g,{\mathbb Y})$ the right action of the group element $g\in\mathcal{G}$ on ${\mathbb Y}$ generated by the left-invariant vector fields, the vielbein transforms as 
\begin{eqnarray}
	{\mathbb E}(\sigma_R(g,{\mathbb Y}))^M&=&g_N{}^M\,{\mathbb E}({\mathbb Y})^N\,, \label{groupact} 
\end{eqnarray}
where $g_N{}^M\in {\rm O}(6,6)$. 
We recall that the adjoint representation of ${\cal G}$ is embedded in O$(6,6)$ as described in (\ref{embedd}). 
For an infinitesimal transformation, parametrized by $\epsilon^M$, the action on ${\mathbb Y}^M$ is described by the corresponding Killing vectors 
\begin{eqnarray}
	{\mathbb Y}^M&\rightarrow &{\mathbb Y}^{\prime M}={\mathbb Y}^M+\delta {\mathbb Y}^M={\mathbb Y}^M+\epsilon^N \,(U^{-1})_N{}^M\,. 
\end{eqnarray}
Its effect on ${\mathbb E}^M$ is the following: 
\begin{eqnarray}
	{\mathbb E}^M({\mathbb Y}^\prime)&=& U({\mathbb Y}^\prime)^M{}_N \,d{\mathbb Y}^{\prime N}={\mathbb E}^M({\mathbb Y})-\epsilon^N\,{\cal T}_{NP}{}^M\,{\mathbb E}({\mathbb Y})^P\,, 
\end{eqnarray}
which is the infinitesimal expansion of (\ref{groupact}). 
An easy way to check this is to contract the Maurer-Cartan equation for ${\mathbb E}^M$ by $\epsilon^M\,\widehat{X}_M$ and use $\iota_{\widehat{X}_M}{\mathbb E}^N=\delta^N_M$: 
\begin{eqnarray}
	\delta {\mathbb E}^M&=&\iota_{\epsilon^N\widehat{X}_N}\,d{\mathbb E}^M=\iota_{\epsilon^N\widehat{X}_N}\,\left(-\frac{1}{2}\,{\cal T}_{IJ}{}^M {\mathbb E}^I\wedge {\mathbb E}^J \right)=-\epsilon^I\,{\cal T}_{IJ}{}^M {\mathbb E}^J\,. 
\end{eqnarray}

The monodromies following by completing the circle on the regular coordinates $y^i \sim y^i +1$ can then be compensated by an action on the $g$ and $B$ fields. 
The twisted reduction Ansatz (\ref{twistedSS}) $$ d{\cal S}^2 = {\cal H}_{MN}(x) \left({\mathbb E}^M + A^M\right)\otimes\left({\mathbb E}^N + A^N \right) $$ shows that an action of a $g \in \Gamma$ on the coordinates ${\mathbb Y} \to \sigma_R(g,{\mathbb Y})$ can either be seen as an action on the doubled vielbeins (\ref{groupact}), or as an action on ${\cal H}_{MN}$ 
\begin{equation}
	{\cal H}_{MN}\left(g_{ij}(x),B_{ij}(x)\right) \to {g_M}^P\, {\cal H}_{PQ}\left(g_{ij}(x),B_{ij}(x)\right) \,{g_N}^Q, \label{reversedaction} 
\end{equation}
and hence on the 4-dimensional fields $g_{ij}$ and $B_{ij}$.

This also provides a criterion for characterizing a compactification as globally geometric, locally geometric or non-geometric. 
We know that the generic action of these monodromies is in O$(6,6)$ and we can split them in the three types of transformations which can be interpreted as a vielbein redefinition (i.e.~coordinate transformation), a $B$-field gauge transformation or a so-called $\beta$ transformation (which includes T-dualities): 
\begin{equation}
	g^M{}_N = \left( 
	\begin{array}{cc}
		A & \beta \\[2mm]
		\Theta & D
	\end{array}
	\right) \label{gact} 
\end{equation}
In order to recover the explicit form of the moduli transformations it is useful to rewrite the action (\ref{reversedaction}) as an action on the matrix 
\begin{equation}
	M_{ij} = g_{ij} + B_{ij}. 
	\label{revmatr} 
\end{equation}
This is realized as the fractional transformation 
\begin{equation}
	M \to (D M + \Theta)(\beta M + A)^{-1}. 
	\label{realM} 
\end{equation}
It is then clear that a $\Theta$ action is simply a gauge transformation $B \to B + \Theta$, while non-geometric transformations mixing $g$ and $B$ follow from any non-vanishing $\beta$. 
So, the group $\mathcal{G}$ defines a geometric compactification if and only if $g$ does not involve $\beta$-transformations. 
This is certainly the case if the matrix $\epsilon^M\,{\cal T}_{MN}{}^P$ has this form, namely if 
\begin{eqnarray}
	\epsilon^M\,{\cal T}_{MN}{}^P&\in&\frak{gl}(n,\mathbb{R})\ltimes \{t^{ij}\}\subset \frak{o}(6,6)\,, \label{form52} 
\end{eqnarray}
where the generators $t^{ij}$ are defined as follows 
\begin{eqnarray}
	t^{ij}&=&\left(\matrix{{\bf 0}& {\bf 0} \cr\delta_{k\ell}^{ij}&{\bf 0}}\right)\,. 
\end{eqnarray}

From the coordinate point of view, we can see that a change of the dual coordinates by a function of the base coordinates corresponds to a gauge transformation of the $B$-field: 
\begin{equation}
	\delta {\mathbb Y}^M = \left\{0,\lambda_i(y)\right\} \Rightarrow \delta B_{ij} = \partial_{[i}\lambda_{j]}(y). 
	\label{coords1} 
\end{equation}
Diffeomorphisms in the base space must be related to similar transformations on the dual space (non-trivial $A$ matrices in (\ref{gact}) acting on both sectors). 
Finally, non-local $\beta$ transformations are related to coordinate transformations
 where the base coordinates are changed by functions of the dual ones 
\begin{equation}
	\delta y^i = f^i(\widetilde y). 
	\label{corsds2} 
\end{equation}
We will see in the next section various examples where these prescriptions are explicitly realised.

In this framework we can also discuss T-duality. 
The most general global transformation of the vielbein basis $\mathbb{E}^M$ which leaves the inner product (\ref{intprod}) has to belong to ${\rm O}(6,6)$: 
\begin{eqnarray}
	\forall \rho\in {\rm O}(6,6)&: & \mathbb{E}^{\prime M}({\mathbb Y}^\prime)=\rho_{N}{}^M\,\mathbb{E}^N({\mathbb Y}^\prime)\,. 
\end{eqnarray}
The new vielbein will now satisfy a different Maurer-Cartan equation 
\begin{eqnarray}
	d\mathbb{E}^{\prime M}&=&X^\prime_{NP}{}^M\,\mathbb{E}^{\prime N}\wedge \mathbb{E}^{\prime P}\,, 
\end{eqnarray}
where 
\begin{eqnarray}
	X^\prime_{NP}{}^M&=&(\rho^{-1})_N{}^I\,(\rho^{-1})_P{}^J\,\rho_K{}^M\,X_{IJ}{}^K\,. 
\end{eqnarray}
The matrix
$\rho$ represents the effect of a T-duality transformation and if it is not in $\mathcal{G}$, the structure of the gauge group will change and thus the properties of the compactification manifold. 
In the next section we shall discuss four T-dual compactifications in detail.

\section{String theory reductions} \label{sec:string_theory_reductions} 

In this section we give several examples where we can apply the general theory described in the previous section. 
First of all we consider one example satisfying the condition (\ref{condRiem}), so that we have a clear description in terms of a metric and B-field, without introducing a non-coordinate basis for the tangent space of the doubled manifold. 
This is the toy model of a non-geometric compactification uncovered in \cite{Kachru:2002sk} through T-dualities\footnote{An interesting alternative interpretation of this model in terms of generalised complex geometry was given in \cite{Petrinitalk}. Its possible relation with the $T + T^*$ splitting was also discussed in \cite{Micu:2007rd}.}. 
Then we will discuss how a non-trivial geometric background, like the flat group compactification of \cite{Scherk:1979zr}, can be described by a non-trivial twisted doubled torus.

\subsection{Fluxes in a doubled 3-torus reduction} 

\label{sub:fluxes_in_a_doubled_3_torus_reduction}

The first model is based on a 3-torus, so that its double is a 6-torus carrying a natural action under the O(3,3) duality symmetry. 
We can always think of this as a special case of a 6-torus where 3 directions are kept fixed. 
The starting point is a flat ${\mathbb T}^3$ with a non-trivial $H$-flux on the global 3-cycle of ${\mathbb T}^3$. 
This has a clear geometric description and can be constructed as a twisted doubled torus by the quotient of a group manifold ${\cal G}$ corresponding to the generalized Heisenberg algebra: 
\begin{equation}
	[\widehat{Z}_i,\widehat{Z}_j] = h_{ijk} \widehat X^k, \quad [\widehat Z_i, \widehat X^j] = 0 = [\widehat X^i, \widehat X^j]. 
	\label{algstart} 
\end{equation}
Then we apply T-duality transformations along the various circles of the original 3-torus showing that the chain of dualities 
\begin{equation}
	h_{xyz} \stackrel{\rho_z}{\longleftrightarrow} \tau^z_{xy}\stackrel{\rho_y}{\longleftrightarrow} Q^{zy}_x \stackrel{\rho_z}{\longleftrightarrow} R^{xyz} \label{chain} 
\end{equation}
works at the level of the gauged supergravity algebra and the corresponding doubled torus formulation.

Duality transformations have a non-trivial action on the vielbeins 
\begin{equation}
	{\mathbb E}^\prime = \rho U({\mathbb Y}^\prime) d{\mathbb Y}^\prime, \quad {\rm for} \quad {\mathbb Y}^\prime = \rho {\mathbb Y}. 
	\label{Etrans} 
\end{equation}
Obviously such an action may transform the vielbeins away from the triangular gauge. 
However, in this example the metric $\gamma$ in (\ref{inhint}) is flat $\gamma = {\cal I}$ and
hence there is always an action $U^\prime = k U$ by a local O(3) $\times$ O(3) transformation $k$ that brings it back to a triangular form. 
Therefore, the new generalized vielbein can always be interpreted in terms of a 3-dimensional vielbein $e_i^a$ and a $B$-field $B_{ij}$. 

A strict application of Busher's rules would require that the generalized metric ${\cal H}$ does not depend on the coordinates corresponding to the directions on which one is acting. 
Therefore, our application of these duality transformations is done at a formal level, but the outcome gives a picture consistent with the expectations. 
The T-dualities along the $z,y$ and $x$ directions of (\ref{chain}) are realized by the following matrices: 
\begin{equation}
	\rho_z = \left( 
	\begin{array}{cccccc}
		1 &0 &0 &0 & 0 &0 \\
		0 & 1 & 0 &0 &0 &0 \\
		0 &0 &0 &0 &0 & 1\\
		0& 0 &0 &1 &0 &0 \\
		0& 0 &0 &0 & 1 &0 \\
		0& 0 & 1 &0 & 0 &0 
	\end{array}
	\right) , \; \rho_y = \left( 
	\begin{array}{cccccc}
		1 & 0 &0 &0 &0 &0 \\
		0 &0 &0 &0 & 1 &0 \\
		0& 0 & 1 &0 & 0 &0 \\
		0& 0 &0 &1 &0 &0 \\
		0& 1 &0 &0 &0 &0 \\
		0& 0 &0 &0 &0 & 1 
	\end{array}
	\right), \; \rho_x = \left( 
	\begin{array}{cccccc}
		0& 0 &0 & 1 &0 &0 \\
		0& 1 &0 &0 &0 &0 \\
		0 &0 & 1 &0 &0 &0 \\
		1 & 0& 0 & 0 & 0 &0 \\
		0 &0 & 0 & 0 & 1 & 0\\
		0 &0 &0 &0 &0 & 1 
	\end{array}
	\right), \label{Tdualities} 
\end{equation}
which belong to the compact subgroup of the full T-duality group.

It is interesting to point out that from the gauged supergravity point of view the compact subgroup of the full duality group is the group of duality transformations that leaves any vacuum of the theory invariant. 
This means that the resulting gauged supergravities give rise to vacua with the same properties, for instance that of having a certain number of moduli fixed. 
However, the higher dimensional interpretation changes. 
This implies that, starting from a vacuum with all moduli stabilized, one can generate other vacua with all moduli stabilized in this way, therefore obtaining new higher-dimensional backgrounds with this same property.

\subsubsection{NS-NS $h$-flux} 

\label{sub:ns_ns_h_flux}

As in \cite{Kachru:2002sk} the starting point is a standard flat 3-torus endowed with constant NS-NS flux\footnote{We should mention that such a background is not a solution of the equations of motion, however it can be promoted to a solution if a non-trivial dilaton is present or if we generalize it to six dimensions and turn on appropriate fluxes.}. 
If $\{x,y,z\}$ are the coordinates on $\mathbb{T}^3$ periodically identified with period $R$, the corresponding doubled torus has coordinates $\mathbb{Y}\equiv\{x,y,z,\tilde x, \tilde y, \tilde z\}$ and $\{\tilde x, \tilde y, \tilde z\}$ are identified with period $1/R$. 
In the following we will take $R = 1$ for simplicity. 
The metric and B-field read 
\begin{equation}
	ds^2 = dx^2+dy^2+dz^2 \quad B=x dy \wedge dz + y dz \wedge dx + z dx \wedge dy 
\end{equation}
and from these data we can immediately write down the generalized vielbein $U_h$ and the corresponding generalized metric ${\cal H}_h = U_h^T U_h$ on the doubled torus 
\begin{equation}
	U_h = \left( 
	\begin{array}{cccccc}
		1 & 0 &0 &0 &0 & 0\\
		0 & 1 &0 &0 &0 &0 \\
		0 & 0 & 1 &0 &0 &0 \\
		0& -z & y &1 & 0 &0 \\
		z &0 & -x & 0 & 1 &0 \\
		-y & x & 0 & 0 &0 & 1 
	\end{array}
	\right), \label{Uh} 
\end{equation}
\begin{equation}
	{\cal H}_h =\left( 
	\begin{array}{cccccc}
		1+y^2+z^2 & -xy & -xz & 0 &z & -y \\
		-xy & 1+x^2+y^2 & -yz &-z & 0 &x \\
		-xz & -yz & 1+x^2+y^2 &y &-x &0 \\
		0 & -z & y &1 &0 &0 \\
		z & 0 & -x & 0 & 1 &0 \\
		-y & x & 0 & 0 &0 & 1 
	\end{array}
	\right) . 
	\label{vielgen1} 
\end{equation}
The generalized coframe reads 
\begin{equation}
	\begin{array}{lclcl}
		E^1 = dx, && E^2 = dy, && E^3 = dz, \\[3mm]
		\widetilde E_1 = d\tilde x - z dy +y dz, && \widetilde E_2 = d\tilde y + z dx - x dz, && \widetilde E_3 = d\tilde z + x dy - y dx, 
	\end{array}
	\label{viel3} 
\end{equation}
and we can verify that they satisfy the following Maurer--Cartan equations 
\begin{equation}
	dE^i = 0, \qquad d\widetilde E_i = -\frac12 \,h_{ijk} E^j E^k, \label{torsions1} 
\end{equation}
with $h_{123} = -2$. 
These structure equations imply that the doubled torus is twisted and that at least locally is a group manifold. 
The corresponding Lie algebra is spanned by a set of generators $\{\widehat Z_i,\widehat X^i\}$ obeying the following commutation relations 
\begin{equation}
	[\widehat Z_i,\widehat Z_j] = h_{ijk} \widehat X^k, \quad [\widehat Z_i,\widehat X^j] =0,\quad [\widehat X^i,\widehat X^j] =0 , 
\end{equation}
which is exactly the gauge algebra of a string compactification on a 3-torus with constant flux $h_{ijk}$. 
The gauge generators $\widehat Z_i$ that originate from the metric are associated with the vielbeins $E^i$ while the dual vielbeins $\widetilde E_i$ correspond to the gauge generators $\widehat X^i$ that come from the reduction of the antisymmetric tensor. 
The $X^i$ generators are central charges and hence they have zero action on the curvature field strengths ($X^i = 0$).

The first observation here is that a different choice of gauge for the $B$-field, for instance $B=3 x dy \wedge dz$, would lead to a generalized coframe that doesn't satisfy the expected Maurer--Cartan equations. 
In particular the structure constants are not fully antisymmetric. 
However, consistency of the gauging implies that the metric ${\cal I}$ is an invariant metric of the gauge algebra and hence the structure constants are always fully antisymmetric. 
The relation between the two choices of $B$ is a gauge transformation $d\Lambda = d(x y)\wedge dz - d(x z) \wedge dy$, which is also an element of O(3,3) 
\begin{equation}
	\Theta = \left( 
	\begin{array}{cccccc}
		1 & & & & & \\
		& 1 & & & & \\
		& & 1 & & & \\
		& z & -y &1 & & \\
		-z & & -2x & & 1 & \\
		y & 2x & & & & 1 
	\end{array}
	\right), \label{theta} 
\end{equation}
that can be applied to the right of (\ref{Uh}) to recover the non-symmetric form of the $B$-field.

A second observation concerns the identifications that make the vielbeins (\ref{viel3}) globally well-defined. 
They are 
\begin{equation}
	\begin{array}{rcl}
		(x,\tilde y,\tilde z) &\sim& (x+1, \tilde y +z, \tilde z -y ),\\[2mm]
		(y,\tilde x,\tilde z) &\sim& (y+1, \tilde x - z, \tilde z +x), \\[2mm]
		(z, \tilde x,\tilde y) &\sim& (z+1, \tilde x +y, \tilde y - x),\\[2mm]
		\tilde x &\sim& \tilde x+1,\\[2mm]
		\tilde y &\sim& \tilde y+1, \\[2mm]
		\tilde z &\sim& \tilde z+1, 
	\end{array}
\end{equation}
and we see that the existence of the non-trivial $B$-field is encoded in the redefinition of the dual coordinates by the actual coordinates, when the latter are shifted. 
In particular, we can interpret the monodromies on the dual coordinates as actions on the effective theory fields and check that this compactification is a well defined geometric compactification according to the discussion of section \ref{sub:global_description}. 
This follows from the fact that the embedding of the gauge generators in O(3,3) is given by $Z_i=- h_{ijk}\,t^{jk}$ and corresponds to a $\Theta$ transformation. 
Actually, we can explicitly see that the action of the monodromy on the $x$, $y$ or $z$ coordinate has to be related to a $B$-field gauge transformation, signalled by the corresponding action on the dual coordinates. 
For instance 
\begin{equation}
	x \sim x+1, \quad \delta \widetilde E_2 = E^3, \quad \delta \widetilde E_3 = -E^2, \label{monexh} 
\end{equation}
can be interpreted as $B_{yz} \to B_{yz} - 1$. 
In this way we recover that when going around the ${\mathbb S}^1$ parameterized by $x$ one has to perform a gauge transformation of the $B$-field to compensate the change in (\ref{monexh}) 
\begin{equation}
	x \sim x + 1, \quad B \sim B - dy \wedge dz. 
	\label{monodh} 
\end{equation}
Similar monodromy relations are obtained for the $y$ and $z$ coordinates.

\subsubsection{Geometric $\tau$ flux} 

\label{sub:geometric_tau_flux}

The first duality transformation is taken along the $z$ direction. 
The new vielbein then is $U_\tau = \rho_z U_h \rho_z$ and it reads 
\begin{equation}
	U_\tau = \left( 
	\begin{array}{cccccc}
		1 & 0 &0 & 0 &0 &0 \\
		0& 1 &0 &0 & 0 & 0\\
		-y& x & 1 & 0 & 0 & 0\\
		0& -\tilde z &0 &1 &0 & y\\
		\tilde z & 0 &0 &0 & 1 & -x\\
		0 & 0 & 0 & 0 &0 &1 
	\end{array}
	\right). 
	\label{vielgen2} 
\end{equation}
The corresponding doubled coframe is given by 
\begin{equation}
	\begin{array}{lclcl}
		E^1 = dx, && E^2 = dy, && E^3 = d z + x dy - y dx, \\[3mm]
		\widetilde E_1 = d\tilde x - \tilde z dy +y d\tilde z, && \widetilde E_2 = d\tilde y + \tilde z dx - x d\tilde z, && \widetilde E_3 = d\tilde z, 
	\end{array}
	\label{viel2} 
\end{equation}
and the associated Maurer--Cartan equations are, as expected, 
\begin{equation}
	dE^i =- \frac12\tau^i_{jk} E^j \wedge E^k, \qquad d \widetilde E_i = \tau_{ij}^k E^j \wedge \widetilde E_k, \label{torsions2} 
\end{equation}
with $\tau_{12}^3 = -2$. 
Again the corresponding Lie algebra matches that obtained from a string theory reduction with geometric flux $\tau^3_{12}$: 
\begin{equation}
	[\widehat Z_i,\widehat Z_j] = \tau_{ij}^k \widehat Z_k, \quad [\widehat Z_i,\widehat X^k] =-\tau_{ij}^k \widehat X^j,\quad [\widehat X^i,\widehat X^j] =0 . 
	\label{algebratau} 
\end{equation}

The coframe is well-defined if we impose the following global identifications 
\begin{equation}
	\begin{array}{rcl}
		(x,\tilde y,z) &\sim& (x+1, \tilde y +z, z -y ),\\[2mm]
		(y,\tilde x,z) &\sim& (y+1, \tilde x - z, z +x), \\[2mm]
		z &\sim& z+1,\\[2mm]
		\tilde x &\sim& \tilde x+1, \\[2mm]
		\tilde y &\sim& \tilde y+1, \\[2mm]
		(\tilde z, \tilde x, \tilde y) &\sim& (\tilde z+1, \tilde x +\tilde y,\tilde y -\tilde x). 
	\end{array}
\end{equation}
In this case the monodromies on the base ``geometric'' coordinates mix them, as expected for an ordinary twisted torus. 
We can actually read from (\ref{viel2}) the corresponding metric and $B$ field 
\begin{equation}
	ds^2=dx^2+dy^2+(dz-ydx+x dy)^2, \quad B=\tilde z dx\wedge dy. 
\end{equation}
When projecting to the base space at $\tilde z = 0$, we see that we obtain a simple globally well defined twisted torus. 
Once more this was expected to be geometric, by inspection of the embedding of the generators in the duality group. 
The monodromies are obtained by $\epsilon^i Z_i + \epsilon_i X^i = \epsilon^i \tau_{ij}^k {t^j}_k + \epsilon_i \tau^i_{jk} t^{jk}$ and they are of the form (\ref{form52}). 
The rotation on the dual coordinate $\tilde z$ has to be done along with a gauge transformation $ B \sim B - dx \wedge dy$. 
In any case everything looks safely geometric.

Also, this background is obviously related to the original T-dual one, but with the non-symmetric $B$-field gauge, by a transformation $\rho_z \Theta \rho_z$. 
This new transformation is not a pure gauge anymore. 
However it leads to the expected purely geometric background without any dependence on the dual coordinates.

\subsubsection{Locally geometric $Q$-flux} 

\label{sub:locally_geometric_q_flux}

The application of a second T-duality along $y$ gives a new generalized vielbein $U_Q=\rho_y U_\tau \rho_y$: 
\begin{equation}
	U_Q = \left( 
	\begin{array}{cccccc}
		1 & 0 & 0 &0 &0 &0 \\
		\tilde z &1 &0 &0 & 0 &-x \\
		-\tilde y& 0 &1 &0 &x & 0\\
		0& 0 &0 &1 & -\tilde z & \tilde y\\
		0 & 0 & 0 &0 &1 & 0\\
		0 &0 & 0 & 0 & 0 &1 
	\end{array}
	\right), \label{vielgen3} 
\end{equation}
and the associated coframes read 
\begin{equation}
	\begin{array}{lclcl}
		E^1 = dx, && E^2 = dy + \tilde z dx - x d\tilde z, && E^3 = d z + x d\tilde y - \tilde y dx, \\[3mm]
		\widetilde E_1 = d\tilde x - \tilde z d\tilde y +\tilde y d\tilde z, && \widetilde E_2 = d\tilde y, && \widetilde E_3 = d\tilde z,\\
	\end{array}
	\label{viel3Q} 
\end{equation}
which satisfy the Maurer--Cartan equations
\begin{equation}
	dE^i =- Q^{ij}_k \widetilde E_j \wedge E^k, \qquad d \widetilde E_i = \frac12 Q^{jk}_i \widetilde E_j \wedge\widetilde E_k, \label{torsions3} 
\end{equation}
with $Q_1^{23} =- 2$. 
This is indeed the correct gauge algebra in the presence of $Q$-flux: 
\begin{equation}
	[\widehat Z_i,\widehat Z_j] = 0 , \quad [\widehat Z_i,\widehat X^j] =-Q^{jk}_i \widehat Z_k,\quad [\widehat X^i,\widehat X^j] =Q^{ij}_k \widehat X^k. 
	\label{Qalgebra} 
\end{equation}

In this case, the doubled-vielbeins are not in the right triangular form, needed to read 
off the actual metric and $B$-field. 
This is achieved by acting on (\ref{vielgen3}) from the left with the following O(3) $\times$ O(3) matrix 
\begin{equation}
	k_2 = \left( 
	\begin{array}{cccccc}
		1 & 0 &0 &0 & 0 &0 \\
		0 & \frac{1}{\sqrt{1+x^2}} & 0 & 0 & 0 & \frac{x}{\sqrt{1+x^2}}\\
		0 & 0 & \frac{1}{\sqrt{1+x^2}} &0 & -\frac{x}{\sqrt{1+x^2}} &0 \\
		0&0 &0 &1 & 0 &0 \\
		0& 0 & \frac{x}{\sqrt{1+x^2}} &0 & \frac{1}{\sqrt{1+x^2}} &0 \\
		0& -\frac{x}{\sqrt{1+x^2}} &0 & 0 & 0 & \frac{1}{\sqrt{1+x^2}} 
	\end{array}
	\right), \label{k2} 
\end{equation}
which leads to 
\begin{equation}
	k_2 U_Q = \left( 
	\begin{array}{cccccc}
		1 &0 &0 &0 & 0 &0 \\
		\frac{\tilde z}{\sqrt{1+x^2}} & \frac{1}{\sqrt{1+x^2}} &0 & 0 & 0 & 0\\
		-\frac{\tilde y}{\sqrt{1+x^2}} & 0 & \frac{1}{\sqrt{1+x^2}} & 0 &0 & 0\\
		0 & 0 &0 &1 & -\tilde z & \tilde y\\[3mm]
		-\frac{x \tilde y}{\sqrt{1+x^2}} &0 & \frac{x}{\sqrt{1+x^2}} &0 & \sqrt{1+x^2} &0 \\
		-\frac{x \tilde z}{\sqrt{1+x^2}} & -\frac{x}{\sqrt{1+x^2}} & 0 &0 & 0 & \sqrt{1+x^2} 
	\end{array}
	\right). 
	\label{vielgen3geo} 
\end{equation}

The identifications that make the space globally defined are 
\begin{equation}
	\begin{array}{rcl}
		(x,y,z) &\sim& (x+1, y +\tilde z, z -\tilde y ),\\[2mm]
		y &\sim& y+1, \\[2mm]
		z &\sim& z+1,\\[2mm]
		\tilde x &\sim& \tilde x+1, \\[2mm]
		(\tilde y,\tilde x,z) &\sim& (\tilde y+1, \tilde x - \tilde z, z +x), \\[2mm]
		(\tilde z, \tilde x, y) &\sim& (\tilde z+1, \tilde x +\tilde y, y - x). 
	\end{array}
\end{equation}
We see here for the first time an identification that shifts the ordinary coordinates by a dual one when identifying a base coordinate. 
This means that when identifying $x \sim x+1$ one has to also identify properly the $B$-field and the metric. 
The action on the vielbeins (\ref{viel3Q}) is indeed 
\begin{equation}
	x \sim x+1, \quad \delta E^2 \sim \widetilde E_3, \quad \delta E^3 \sim- \widetilde E_2 ,\label{actv} 
\end{equation}
and this is a $\beta$ transformation of the form 
\begin{equation}
	\beta = \left( 
	\begin{array}{cccccc}
		1 &&&& & \\
		&1&&&& \\
		&&1&&&\\
		& & &1&& \\
		& & -1 &&1&\\
		&  1 & &&&1 
	\end{array}
	\right). 
	\label{formbeta} 
\end{equation}
As explained in section \ref{sub:global_description} we can read the transformation required on the moduli fields by the action on the matrix $M = g + B$. 
The result is that 
\begin{equation}
	\begin{array}{l}
		\begin{array}{rcl}
			g_{xx} &\to& \frac{1}{\Delta} \left(g_{xx}(1 - B_{yz})^2 + 2 B_{xz} g_{xy} - 2 B_{xz} B_{yz} g_{xy} - 2 B_{xy} g_{xz} + 2 B_{xy} B_{yz} g_{xz} \right.\\[2mm]
			&&\left. 
			+ B_{xz}^2 g_{yy} - g_{xz}^2 g_{yy} - 2 B_{xy} B_{xz} g_{yz} + 2 g_{xy} g_{xz} g_{yz} - g_{xx} g_{yz}^2 + B_{xy}^2 g_{zz} \right.\\[2mm]
			&&\left.- g_{xy}^2 g_{zz} + g_{xx} g_{yy} g_{zz}\right) , 
		\end{array}
		\\[12mm]
		\begin{array}{rclcrcl}
			g_{xy} &\to&\displaystyle \frac{g_{xy}(1-B_{yz})-B_{xy}g_{yz}+B_{xz}g_{yy}}{\Delta}, && B_{xy} &\to& \displaystyle\frac{B_{xy}(1-B_{yz})-g_{xy}g_{yz}+g_{xz}g_{yy}}{\Delta},\\[3mm]
			g_{xz} &\to&\displaystyle \frac{g_{xz}(1-B_{yz})+B_{xz}g_{yz}-B_{xy}g_{zz}}{\Delta},&& B_{xz} &\to&\displaystyle \frac{B_{xz}(1-B_{yz})+g_{xz}g_{yz}-g_{xy}g_{zz}}{\Delta},\\[3mm]
			g_{yz} &\to& \displaystyle\frac{g_{yz}}{\Delta}, && B_{yz} &\to& \displaystyle\frac{-B_{yz}(1-B_{yz})-g_{yz}^2+g_{yy}g_{zz}}{\Delta},\\[3mm]
			g_{yy} &\to&\displaystyle \frac{g_{yy}}{\Delta}, && g_{zz} &\to&\displaystyle \frac{g_{zz}}{\Delta} , 
		\end{array}
	\end{array}
	\label{trasfo} 
\end{equation}
with $\Delta = - g_{yz}^2+(1- B_{yz})^2 + g_{yy}g_{zz}$. 
This means that the metric and $B$-field give a proper geometric description to the base space only if we act by (\ref{trasfo}) whenever $x \sim x +1$. 
This is also clear when looking at the metric and $B$-field after projecting to the base and setting the dual coordinates to zero: 
\begin{equation}
	ds^2=dx^2+\frac{1}{1+x^2}(dy^2+dz^2),\quad B=\frac{x}{1+x^2} dy\wedge dz. 
	\label{Qspace} 
\end{equation}
Both the metric and $B$-field are not well defined functions of the $x$ coordinate, which is periodically identified. 
They give however a good global description upon using the identification (\ref{trasfo}), which for this case reduces to \cite{Kachru:2002sk, Lowe:2003qy}: 
\begin{equation}
	\begin{array}{rclcrcl}
		g_{xx} &\to&\displaystyle g_{xx}&& B_{yz} &\to& \displaystyle\frac{-B_{yz}(1-B_{yz})+g_{yy}g_{zz}}{\Delta},\\[3mm]
		g_{yy} &\to&\displaystyle \frac{g_{yy}}{\Delta} && g_{zz} &\to&\displaystyle \frac{g_{zz}}{\Delta} ,
	\end{array}
	\label{trasfored} 
\end{equation}
with $\Delta = (1- B_{yz})^2 + g_{yy}g_{zz}$, because $g_{xy} = g_{xz} = g_{yz} = B_{xy} = B_{xz} = 0$.

Once more, a different way to obtain (\ref{Qspace}) without any dependence on the dual coordinates is by taking the action of the duality transformations directly on the non-symmetric gauge for the original $B$-field. 
This is dual to the above background by a $\rho_y \rho_z \Theta \rho_z \rho_y$ transformation.

\subsubsection{Non-geometric $R$-flux} 

\label{sub:non_geometric_r_flux}

The third T-duality at our disposal yields the generalized metric 
\begin{equation}
	U_R = \left( 
	\begin{array}{cccccc}
		1 & 0 & 0 &0 &-\tilde z &\tilde y \\
		0 &1 &0 &\tilde z & 0 &-\tilde x \\
		0& 0 &1 &-\tilde y &\tilde x & 0\\
		0& 0 & 0&1 & 0 &0\\
		0 &0 & 0 &0 &1 & 0\\
		0 &0 & 0 & 0 & 0 &1 
	\end{array}
	\right), \label{vielgen4} 
\end{equation}
whose corresponding coframes are 
\begin{equation}
	\begin{array}{rclcrclcrcl}
		E^1 & = & dx +\tilde y d\tilde z - \tilde z d\tilde y, &\quad & E^2 & = & dy + \tilde z d \tilde x -\tilde x d\tilde z, &\quad& E^3 & = & d z + \tilde x d\tilde y - \tilde y d\tilde x, \\[3mm]
		\widetilde E_1 & = & d\tilde x, &\quad& \widetilde E_2 & = & d\tilde y, &\quad& \widetilde E_3 & = & d\tilde z,\\
	\end{array}
	\label{viel4} 
\end{equation}
satisfying the Maurer--Cartan equations 
\begin{equation}
	dE^i =-\frac12\, R^{ijk} \widetilde E_j \wedge \widetilde E_k, \qquad d \widetilde E_i = 0. 
	\label{torsions4} 
\end{equation}
Therefore the doubled torus indeed realized the gauge algebra with $R$-flux: 
\begin{equation}
	[\widehat Z_i,\widehat Z_j] = 0 , \quad [\widehat Z_i,\widehat X^j] =0,\quad [\widehat X^i,\widehat X^j] = R^{ijk} \widehat Z^k. 
	\label{Ralgebra} 
\end{equation}

A further action on the left with the following O(3) $\times$ O(3) transformation 
\begin{equation}
	k_3 = \chi \left( 
	\begin{array}{cccccc}
		1 + \tilde x^2& \tilde x \tilde y + \tilde z& -\tilde y + \tilde x \tilde z& -\tilde y^2 - \tilde z^2 & \tilde x \tilde y + \tilde z & -\tilde y + \tilde x \tilde z \\
		\tilde x \tilde y - \tilde z & 1+ \tilde y^2 & \tilde x + \tilde y \tilde z & \tilde x \tilde y - \tilde z & -\tilde x^2 - \tilde z^2 & \tilde x + \tilde y \tilde z \\
		\tilde y + \tilde x \tilde z & -\tilde x + \tilde y \tilde z & 1+ \tilde z^2 & \tilde y + \tilde x \tilde z & -\tilde x + \tilde y \tilde z & -\tilde x^2 - \tilde y^2 \\
		-\tilde y^2 - \tilde z^2 & \tilde x \tilde y + \tilde z & -\tilde y + \tilde x \tilde z & 1+ \tilde x^2 & \tilde x \tilde y + \tilde z & -\tilde y + \tilde x \tilde z\\
		\tilde x \tilde y - \tilde z & -\tilde x^2 - \tilde z^2 & \tilde x + \tilde y \tilde z & \tilde x \tilde y - \tilde z & 1 + \tilde y^2 & \tilde x + \tilde y \tilde z\\
		\tilde y + \tilde x \tilde z & -\tilde x + \tilde y \tilde z & -\tilde x^2 - \tilde y^2 & \tilde y + \tilde x \tilde z & -\tilde x +\tilde y \tilde z & 1 +\tilde z^2 
	\end{array}
	\right) \label{k3} 
\end{equation}
where $\chi = \frac{1}{1+\tilde x^2+\tilde y^2 + \tilde z^2}$ brings the vielbein into a triangular form: 
\begin{equation}
	k_3 g_x g_y g_z U_R = \left( 
	\begin{array}{cccccc}
		\chi (1+\tilde x^2) & \chi (\tilde x \tilde y +\tilde z) & \chi (-\tilde y + \tilde x \tilde z) &0 &0 &0 \\
		\chi (\tilde x\tilde y-\tilde z) &\chi (1+\tilde y^2) &\chi (\tilde x+\tilde y \tilde z) &0 & 0 &0 \\
		\chi (\tilde y + \tilde x \tilde z)& \chi (-\tilde x + \tilde y\tilde z) &\chi (1+\tilde z^2) &0 &0 & 0\\
		-\chi (\tilde y^2 +\tilde z^2) & \chi (\tilde x\tilde y + \tilde z) &\chi (-\tilde y + \tilde x\tilde z) &1 & \tilde z &-\tilde y\\
		\chi (\tilde x\tilde y -\tilde z) &-\chi (\tilde x^2+\tilde z^2) & \chi (\tilde x+\tilde y \tilde z) &-\tilde z &1 & \tilde x\\
		\chi (\tilde y + \tilde x \tilde z) &\chi (-\tilde x+\tilde y \tilde z) &- \chi (\tilde x^2 +\tilde y^2) & \tilde y & -\tilde x &1 
	\end{array}
	\right). 
	\label{vielgen5} 
\end{equation}
The global identifications of this space are 
\begin{equation}
	\begin{array}{rcl}
		x &\sim& x+1, \\[2mm]
		y &\sim& y+1, \\[2mm]
		z &\sim& z+1,\\[2mm]
		(\tilde x,y,z) &\sim& (\tilde x+1, y +\tilde z, z -\tilde y ),\\[2mm]
		(\tilde y,x,z) &\sim& (\tilde y+1, x - \tilde z, z +\tilde x), \\[2mm]
		(\tilde z,x, y) &\sim& (\tilde z+1, x +\tilde y, y -\tilde x). 
	\end{array}
\end{equation}
Although the naive projection to the base space may seem to yield a flat torus with a trivial $B$-field, the identifications required on the dual coordinates have an extreme effect on the field content. 
These identifications involve $\beta$-transformations related to the shift of a base coordinate by the dual ones. 
If we insist on interpreting these identifications as actions on the space-time fields we obtain a fully non-geometric background, because one has to perform identifications that entangle the metric and the $B$-field, without any relation to a geometric action on the base coordinates. 
It is actually known that this space is isomorphic to a so-called non-associative torus \cite{Bouwknegt:2004ap, Ellwood:2006my, Grange:2006es}, which does not admit a classical geometric description even locally.

\subsubsection{A summary} 

All previous examples can be grouped in a unique class of compactifications on a 3-torus with general flux in which the gauge generators in the adjoint representation are nilpotent of order two: $X_I\cdot X_J=0$. 
The doubled torus has vielbein satisfying eq.~(\ref{torsions}) with $\mathcal{T}_{IJ}{}^K=-X_{IJ}{}^K$. 
The previous four cases correspond to flux tensors $X_{NM}{}^P$ identified with $H_{ijk},\,\tau_{ij}{}^k,\,Q_i{}^{jk}$ and $R^{ijk}$ related to one another by T-duality, respectively.

These vielbeins will follow from the left-invariant one-form 
\begin{eqnarray}
	g^{-1}dg&=&U^J{}_I\,d\mathbb{Y}^I\,\widehat{X}_J, 
\end{eqnarray}
but it is also useful to define the right-invariant ones
\begin{equation}
	dg\,g^{-1}=\widetilde{U}^J{}_I\,d\mathbb{Y}^I\,\widehat{X}_J ,
\end{equation}
where, due to the nilpotency of $X_I$, the matrices $U$ and $\tilde{U}$ are $\mathcal{G}$ elements and have the form 
\begin{eqnarray}
	U^J{}_I&=&\delta^J_I+\frac{1}{2}\,y^M\,X_{MI}{}^J=\exp\left(\frac{1}{2}\,y^M\,X_M\right)^J{}_I\,,\nonumber\\
	\widetilde{U}^J{}_I&=&\delta^J_I-\frac{1}{2}\,y^M\,X_{MI}{}^J=\exp\left(-\frac{1}{2}\,y^M\,X_M\right)^J{}_I\,. 
\end{eqnarray}
In terms of the above matrices we can write the infinitesimal variations of $\mathbb{Y}^M$ under the left and right action of $\mathcal{G}$: 
\begin{eqnarray}
	\mbox{Left action:}&&\mathbb{Y}^{\prime M}= \mathbb{Y}^M+\epsilon_{(L)}^N\,\widetilde{U}^{-1}_N{}^M=\mathbb{Y}^M+\epsilon^M_{(L)}+\frac{1}{2}\,\epsilon_{(L)}^N\,\mathbb{Y}^IX_{IN}{}^M\,,\label{lact}\\
	\mbox{Right action:}&&\mathbb{Y}^{\prime M}= \mathbb{Y}^M+\epsilon_{(R)}^N\,{U}^{-1}_N{}^M=\mathbb{Y}^M+\epsilon^M_{(R)}-\frac{1}{2}\,\epsilon_{(R)}^N\,\mathbb{Y}^IX_{IN}{}^M\,.\label{ract} 
\end{eqnarray}
In virtue of the nilpotency of $X_I$, the above transformation rules hold also for finite $\epsilon^M$. 
Therefore we can use eq.~(\ref{lact}) for integer $\epsilon_{(L)}^M=n^M$, to define the action of $\Gamma$: 
\begin{eqnarray}
	\mathbb{Y}^M&\sim&\mathbb{Y}^M+n^M+\frac{1}{2}\,n^N\,\mathbb{Y}^IX_{IN}{}^M\,,\label{gacte} 
\end{eqnarray}
which will define our left quotient $\Gamma\backslash{\cal G}$. 
If we perform now a simultaneous left and right action with constant integer parameters $n=\epsilon_{(R)}=\epsilon_{(L)}$, the effect is to independently shift each coordinate: 
\begin{eqnarray}
	\mbox{Left/right diagonal action:}&&\mathbb{Y}^{\prime M}=\mathbb{Y}^M+ n^M\,.\label{lract} 
\end{eqnarray}
Howeveer, the vielbein $\mathbb{E}^M$ will feel only the right component of the transformation: 
\begin{eqnarray}
	\mathbb{E}(\mathbb{Y}^\prime)^M&=&s_N{}^M\,\mathbb{E}(\mathbb{Y})^N=\mathbb{E}(\mathbb{Y})^N-\frac{1}{2}\,n^N\,X_{NI}{}^M\,\mathbb{E}(\mathbb{Y})^I\,,\label{monos} 
\end{eqnarray}
The matrix $s$ has the general form (\ref{gacte}) and it can be absorbed in a corresponding duality of the four dimensional $g_{ij}$ and $B_{ij}$ moduli according to (\ref{realM}). 
This led to the monodromies discussed above for the four cases in which $X_{MN}{}^P=\,h_{ijk},\,\tau_{ij}{}^k,\,\,Q_i{}^{jk},\,\,R^{ijk}$. 
From equation (\ref{monos}) we see that only in the two latter cases a shift (\ref{lract}) in the coordinates may involve a $\beta$-transformations. 
In the presence of $Q$-flux the $\beta$ transformation may be induced by a shift $y^i\rightarrow y^i+n^i$ in the $y^i$ coordinates and $\beta^{ij}=n^k\,Q_{k}{}^{ij}$, while in the presence of $R$-flux a $\beta$-transformation may be induced by a shift $\tilde{y}_i\rightarrow \tilde{y}_i+n_i$: $\beta^{ij}=n_k\,R^{kij}$. 

\subsection{The flat group} 

\label{sub:the_flat_group}

Another interesting example is given by the so-called flat groups. 
These group manifolds were introduced in \cite{Scherk:1979zr} as means of generating a potential admitting a $D$-dimensional Minkowski vacuum upon reducing a $D+n$-dimensional theory on a ${\mathbb T}^n$ torus. 
This compactification is equivalent to performing first an ordinary reduction from $D+n$ to $D+1$ dimensions and then a Scherk--Schwarz reduction on the last compactification circle with a twist that depends on its coordinate and on a matrix $M\in \frak{so}(n-1)$. 
Splitting the index running on the $n$ extra coordinates $i=0,\dots, n-1$ into $a=1,\dots, n-1$ and $0$, we can write the resulting gauge algebra as 
\begin{eqnarray}
	[Z_0,\,Z_a]&=&M_a{}^b\,Z_b. 
\end{eqnarray}
When dealing with the full reduction of a supergravity theory and not just with the gravity sector one has additional generators $X^i$ corresponding to the gauge vectors coming from the reduction of tensor fields in higher dimensions. 
Altogether these generators describe the gauge algebra 
\begin{eqnarray}
	[\widehat Z_0,\,\widehat Z_a]&=&M_a{}^b\,\widehat Z_b\,\,;\,\,\,[\widehat Z_0,\,\widehat X^a]=-M_b{}^a\,\widehat X^b\,\,;\,\,\,[\widehat X^a\,,\widehat Z_b]=-M_b{}^a\, \widehat X^0\,, \label{faithflat} 
\end{eqnarray}
which is realized as a faithful representation on the gauge vector fields of the reduced theory: 
\begin{equation}
	\begin{array}{rcl}
		\delta A^0_\mu &=& \partial_\mu \Lambda^0, \\[2mm]
		\delta A^a_\mu &=& \partial_\mu \Lambda^a + M_b{}^a \Lambda^0 A_\mu^b- M_b{}^a \Lambda^b A_\mu^0, \\[2mm]
		\delta A_{\mu a} &=& \partial_\mu \Lambda_a+ M_a{}^b \Lambda_0 A_{\mu b}- M_a{}^b \Lambda_b A_{\mu 0}, \\[2mm]
		\delta A_{\mu 0} &=& \partial_\mu \Lambda_0+ M_a{}^b \Lambda^a A_{\mu b}- M_a{}^b \Lambda_b A^a_{\mu}. 
	\end{array}
	\label{vectoraction} 
\end{equation}
The role of $\widehat X^0$ is that of a central charge and therefore we see that in this example there is an abelian ideal which should be removed when embedding (\ref{faithflat}) inside the algebra corresponding to the duality group. 
The algebra (\ref{faithflat}) is not a subalgebra of the duality one, namely $\frak{o}(n,n)$. 
However, the adjoint action of $X^0$ is trivial on all the curvatures and therefore the adjoint representation can be embedded in $\frak{o}(n,n)$, whose generators we name $t_i{}^j$ for the $ \frak{sl}(n,\mathbb{R})$ part and $t^{ij},\,t_{ij}$ for the so-called $B$ and $\beta$ transformations respectively. 
Writing the $\frak{o}(n,n)$ algebra as 
\begin{eqnarray}
	[t_i{}^j,\,t_k{}^\ell]&=&\delta^j_k\,t_i{}^\ell-\delta^\ell_i\,t_k{}^j\,,\nonumber\\
	\left[ t_i{}^j,\, t^{k\ell}\right]&=& 2\,\delta^{[k}_i\,t^{\ell]j}\,, 
\end{eqnarray}
the gauge generators in the adjoint representation are obtained by 
\begin{eqnarray}
	Z_0&=&M_a{}^b\,t_b{}^a\,\,;\,\,\,Z_a=M_a{}^b\,t_b{}^0\,\,;\,\,\,X^a=M_b{}^a\,t^{0b}\,. 
\end{eqnarray}
The resulting algebra is 
\begin{eqnarray}
	[Z_0,\,Z_a]&=&M_a{}^b\,Z_b\,\,;\,\,\,[Z_0,\,X^a]=-M_b{}^a\,X^b\,\,;\,\,\,[X^a\,,Z_b]=0\, 
\end{eqnarray}
and it is a subalgebra of $\frak{o}(n,n)$ by construction.

To construct the corresponding doubled torus, one can start from a group representative 
\begin{eqnarray}
	g&=&\exp(\tilde{y}_0\,\widehat X^0 )\,\exp(\tilde{y}_a\,\widehat X^a)\,\exp(y^a\,\widehat Z_a)\,\exp(y^0\,\widehat Z_0),
\end{eqnarray}
and obtain the left-invariant vielbeins from 
\begin{eqnarray}
	g^{-1}dg&=&(d\tilde{y}_0+y^a\,d\tilde{y}_b\,M_a{}^b)\,\widehat X^0+d\tilde{y}_b\,[\exp(y^0\,M)]_a{}^b\,\widehat X^b+\nonumber\\&&+dy^b\,[\exp(-y^0\,M)]_b{}^a\,\widehat Z_a+dy^0\,\widehat{Z}_0=d{\mathbb Y}^M\,U_M{}^N\,\widehat{X}_N\,. 
\end{eqnarray}
The matrix $U$ reads 
\begin{eqnarray}
	U&=&\left(\matrix{1&0&0&0\cr 0& [\exp(-y^0\,M)]_a{}^b &0&0\cr 0&0&1&0 \cr 0&0& y^c\,M_c{}^a & [\exp(y^0\,M)]_b{}^a}\right)\,. 
\end{eqnarray}
As expected from the previous general discussion, the matrix $U^N{}_M$ is not an O$(n,n)$ matrix and it cannot be put in the triangular form described in (\ref{triangular}). 
However, as noticed in section \ref{sub:ordinary_geometry_from_a_doubled_space}, one can define the dual basis to the cotangent space described by $\{dy^0, dy^a\}$ by defining: 
\begin{equation}
	\frac{\partial\phantom{y}}{\partial y^0} \equiv d \widetilde y_0 + y^c M_c{}^a d \widetilde y_a, \quad \frac{\partial\phantom{y}}{\partial y^a} \equiv d\widetilde y_a. 
	\label{defcot} 
\end{equation}
By using the $\{dy^i, \partial_i\}$ basis the vielbein ${\mathbb E}^M$ can be described in terms of an O$(n,n)$ matrix which in this case is already in triangular form. 
It describes the vielbeins of the twisted torus geometry for the flat group and a zero $B$-field. 
For instance, for a simple 3-torus one has 
\begin{equation}
	\left\{ 
	\begin{array}{rcl}
		E^1 &=& \cos y^0 \, dy^1 + \sin y^0 \, dy^2, \\[2mm]
		E^2 &=& -\sin y^0 \, dy^1 + \cos y^0 \, dy^2, \\[2mm]
		E^0 & = & dy^0, \\[2mm]
		\widetilde E_1 &=& \cos y^0 \, d \widetilde y_1 + \sin y^0 \, d \widetilde y_2, \\[2mm]
		\widetilde E_2 &=& -\sin y^0 \, d \widetilde y_1 + \cos y^0 \, d \widetilde y_2,\\[2mm]
		\widetilde E_3 &=& d \widetilde y_0 - y^2 d \widetilde y_1 + y^1 d \widetilde y_2 
	\end{array}
	\right. 
	\label{vielbflat} 
\end{equation}
and, after identifying $\partial_1 = d\widetilde y_1$, $\partial_2 = d\widetilde y_2$ and $\partial_0 = d\widetilde y_0 -y^2 d \widetilde y_1 + y^1 d \widetilde y_2$, it reduces to a triangular form. 
This background is definitely geometric and indeed, in the twisted doubled torus description, if we transform ${\mathbb Y}^M$ by means of an infinitesimal $\mathcal{G}$ transformation 
\begin{eqnarray}
	\delta y^0 &=&\epsilon^0 ,\nonumber\\
	\delta y^a &=&\epsilon^b\,(e^{y^0 \,M})_b{}^a,\nonumber\\
	\delta y_0 &=&\epsilon_0-\epsilon_a\,y^b\,(M\,e^{-y^0 \,M})_b{}^a,\nonumber\\
	\delta y_a &=&\epsilon_b\,(e^{-y^0 \,M})_a{}^b,\nonumber 
\end{eqnarray}
the vielbein forms ${\mathbb E}^M$ transform according to 
\begin{eqnarray}
	\epsilon^M\,T_{MN}{}^P&=&\left( \matrix{ 0&0&0&0\cr -\epsilon^c\,M_c{}^b& \epsilon^0\,M_a{}^b &0&0\cr 0&-\epsilon_c\,M_b{}^c&0&\epsilon^c\,M_c{}^b \cr \epsilon_c\,M_b{}^c&0& 0& -\epsilon^0\,M_b{}^a }\right)\,, 
\end{eqnarray}
which is in the geometric subgroup.

\subsection{${\cal N}=4$ gaugings and non-geometric fluxes}

The set of fluxes considered in this section, namely the physical NS-NS $h$-flux, the geometric $\tau$-flux and their non-geometric counterparts $Q$ and $R$, appear naturally when one considers toroidal compactifications of the common sector consisting of the metric and the $B$ field. 
Therefore, it makes sense to focus on heterotic string theory whose massless spectrum contains no other higher rank forms (it also includes gauge fields but we neglect them for simplicity). 
A reduction of heterotic supergravity on a $\mathbb{T}^6$ with any of the above fluxes turned-on yields a four-dimensional theory with 16 supercharges and non-abelian gauge fields, i.e.~an ${\cal N}=4$ gauged supergravity.

These theories have been recently constructed in full generality \cite{Schon:2006kz}. 
They are characterized by two types of embedding tensors: $f_{\alpha IJK}$ and $\xi_{\alpha I}$, where the index $\alpha$ denotes a doublet under the SL$(2, \mathbb{R})$ factor of the duality group and $I,J,K$ are in the fundamental of O(6,6) and completely antisymmetrized. 
Notice that we consider a reduction only of the gravity sector, therefore we obtain 12 gauge fields rotated by the O(6,6) duality group. 
Reducing the gauge fields already present in 10 dimensions would result in more four-dimensional gauge fields and an appropriate enlargement of the duality group. 

It is convenient to set the embedding tensors $f_{-IJK}$ and $\xi_{-I}$ to zero, because one usually is interested in electric gaugings, i.e.~gaugings where only the electric gauge potentials have non-abelian interactions. 
Splitting the fundamental indices of O(6,6) as $I={{}^i,{}_i}$ with $i=1,2,\ldots,6$ and so that the metric (\ref{invmetric}) takes the form ${\cal I}_{ij}={\cal I}^{ij}=0, {\cal I}_{i}{}^j={\cal I}^j{}_{i}=\delta_{i}^j$, enables us to separate the embedding tensor $f_{+IJK}$ into four types: 
\begin{equation}
	f_{+ijk} \sim h_{ijk},\quad f_{+ ij}{}^k \sim \tau^{k}_{ij},\quad f_{+}{}^{ij}{}_k \sim Q^{ij}_{k}, \quad f_{+}{}^{ijk} \sim R^{ijk}. 
\end{equation}
These
different types of gauging parameters correspond to the set of fluxes under consideration\footnote{The higher-dimensional origin of the other embedding tensor $\xi_{+I}$ was elucidated in \cite{Derendinger:2007xp} and corresponds to a reduction with a duality twist inside the O(1,1) $\subset {\rm SL}(2,\mathbb{R})$ part of the full duality group.}. 
This correspondence can be easily checked by matching the potentials of ${\cal N}=4$ gauged supergravity with those obtained from direct reductions of heterotic supergravity in the case of physical and/or geometric fluxes and with their extensions, motivated by duality arguments, in the case of non-geometric fluxes \cite{Shelton:2005cf,Aldazabal:2006up}.

\section{M-theory reductions} \label{sec:m_theory_reductions} 

In the following we shall try to extend our analysis to M-theory reductions to four dimensions. 
It is known that the effective theory describing the low-energy dynamics of M-theory (eleven dimensional supergravity \cite{Cremmer:1978km}) on a seven-torus $\mathbb{T}^7$ is an ungauged ${\cal N}=8,\,D=4$ supergravity. 
It was shown in \cite{Cremmer:1979up} that the manifest ${\rm GL}(7,\mathbb{R})$ global symmetry of the four dimensional theory, associated with the $\mathbb{T}^7$-compactification, is enhanced to an ${\rm E}_{7(7)}$ global symmetry of the equations of motion and Bianchi identities once the seven 2-forms, arising from the reduction of the 3-form, are dualized to scalar fields. 
In this framework the duality group ${\rm E}_{7(7)}$ plays the role of the group ${\rm SL}(2,\mathbb{R})\times {\rm O}(6,6)$ in the heterotic case (or of the group ${\rm O}(1,1)\times {\rm O}(6,6)$ if the 2-form is not dualized to a scalar). 
In contrast to this case, in which the duality action of the ${\rm O}(6,6)$ group on the vector field strengths and their magnetic duals is block-diagonal, namely it does not mix electric with magnetic charges, the duality action of ${\rm E}_{7(7)}$ is non-perturbative. 
In fact the electric and magnetic charges transform all together in the representation ${\bf 56}$ of ${\rm E}_{7(7)}$. 

The eleven dimensional origin of the various four-dimensional fields can be recovered by branching the corresponding ${\rm E}_{7(7)}$ representations with respect to ${\rm GL}(7,\mathbb{R})\subset {\rm E}_{7(7)}$. 
For instance, the branching 
\begin{eqnarray}
	{\bf 56}&\rightarrow & {{\bf 7}}^\prime_{-3}+{\bf 21}_{-1}+{\bf 7}_{+3}+ {{\bf 21}}^\prime_{+1}\,,\label{branch56} 
\end{eqnarray}
allows us to identify the ${{\bf 7}}^\prime_{-3}$ with the Kaluza-Klein (KK) vectors $A^i_\mu$, ($i=1,\dots, 7$), the ${\bf 21}_{-1}$ with the vectors $A_{ij\mu}$ originating from the eleven dimensional 3-form, and the remaining representations with the corresponding magnetic dual vector potentials. 
The electric and magnetic charges also split according to (\ref{branch56}) into $\{p^i,\,p_{ij},\,q_i,\,q^{ij}\}$, where $q_i$ are the quantized momenta, $p^i$ are the KK monopole charges while $q^{ij}$ and $\epsilon^{i_1\dots i_5\,ij}\,p_{ij}$ are the charges of $M2$ and $M5$-branes wrapped along the cycles $(i,j)$ and $(i_1\dots i_5)$ of $\mathbb{T}^7$, respectively.

In this setup, the presence of fluxes induces local symmetries in the four-dimensional theory, which can thus be constructed from the ungauged $D=4,\,{\cal N}=8$ theory by \emph{gauging} a suitable Lie group ${\cal G}$. 
The most general gauging in the maximal four dimensional theory was discussed in \cite{deWit:2005ub,de Wit:2007mt}. 
It was shown that the gauged field equations and Bianchi identities can be written formally in a ${\rm E}_{7(7)}$ invariant way. 
This was done by gauging $56$ gauge generators $ X_M$, $M=1,\dots, 56$, in $\frak{e}_{7(7)}$ by means of all the $56$ vector fields $A^M_\mu=(A^\Lambda_\mu,\,A_{\Lambda\mu})$, $\Lambda=1,\dots, 28$, (which include the magnetic potentials $\,A_{\Lambda\mu}$). 
The adjoint representation of the $X_M$ generators is required to be in the ${\bf 56}$ of $\frak{e}_{7(7)}$ and can be expanded in a basis $\{t_\alpha\}$ of $\frak{e}_{7(7)}$ generators through an embedding tensor 
\begin{eqnarray}
	X_M&=&\theta_M{}^\alpha\,t_\alpha\,. 
\end{eqnarray}
It is useful to define the tensor $X_{MN}{}^P$ as the matrix representation of $X_M$ in the ${\bf 56}$: $X_{MN}{}^P=\theta_M{}^\alpha\,t_{\alpha\,N}{}^P$. 
Since the representation ${\bf 56}$ is symplectic, one can use the symplectic invariant matrix $\Omega^{MN}$ ($\Omega_{MN}$) to raise (lower) indices. 
Thus if we denote by $d_{\alpha \,MN}=t_{\alpha\,M}{}^P\,\Omega_{PN}$, the invariance of $\Omega$ under $t_\alpha$ implies $d_{\alpha \,MN}=d_{\alpha \,NM}$. 
The gauge generators close a $56$ dimensional gauge algebra with structure: 
\begin{eqnarray}
	[ X_M,\, X_N]&=&\mathcal{T}_{MN}{}^P\, X_P=-X_{MN}{}^P\, X_P\,.\label{e7gauge} 
\end{eqnarray}

The most general deformation of the ${\cal N}=8,\,D=4$ theory is then encoded in the ${\rm E}_{7(7)}$ covariant tensor $\theta_M{}^\alpha$. 
Consistency of the gauging with ${\cal N}=8$ supersymmetry requires $\theta_M{}^\alpha$ (or equivalently $X_{MN}{}^P$) to transform in the representation ${\bf 912}$ of ${\rm E}_{7(7)}$. 
This linear condition can be expressed in the form $X_{(MNP)}=0$. 
Besides the linear one, $\theta_M{}^\alpha$ is also subject to the quadratic constraint (\ref{e7gauge}), which expresses the closure of the gauge algebra inside $\frak{e}_{7(7)}$ or, equivalently, the gauge invariance of the embedding tensor itself, and which can be recast in the following form: 
\begin{eqnarray}
	X_{MN}{}^P\,X_{RP}{}^Q-X_{RN}{}^P\,X_{MP}{}^Q+X_{MR}{}^P\,X_{PN}{}^Q&=&0\,,\label{2con} 
\end{eqnarray}
or equivalently, using the linear constraint, as 
\begin{eqnarray}
	\Omega^{MN}\,\theta_M{}^\alpha\,\,\theta_N{}^\beta&=&0\,.\label{2con2} 
\end{eqnarray}

The above condition guarantees that no more than 28 vector fields take part into the minimal couplings and thus prevents the theory, which involves magnetic couplings as well as electric ones, to suffer from locality problems. 
It is important to notice that the tensor $X_{MN}{}^P$ is not antisymmetric in the first two indices and thus it is not proportional to the structure constants. 
Group theoretical arguments show that the tensors $X$ and $\mathcal{T}$ are proportional only when contracted with a gauge generator, as in eq.~(\ref{e7gauge}). 
As a consequence, eq.~(\ref{2con}) does not imply the Jacobi identity for $X_{MN}{}^P$. 
The closest we can get to it is through the following identity, which can be derived using (\ref{2con}): 
\begin{eqnarray}
	X_{[MN]}{}^P\,X_{[RP]}{}^Q+X_{[NR]}{}^P\,X_{[MP]}{}^Q+X_{[RM]}{}^P\,X_{[NP]}{}^Q&=&\frac{1}{2}\,X^Q{}_{P[R}\,X_{MN]}{}^P\,.\label{2con3} 
\end{eqnarray}
In virtue of eq.~(\ref{2con2}), the right hand side of the above identity vanishes upon contraction with $X_Q$ and thus that the Jacobi identity holds for the commutation relation (\ref{e7gauge}). 

The construction in \cite{de Wit:2007mt} also requires the introduction of $133$ tensor fields $B_{\mu\nu\alpha}$ transforming in the adjoint representation of ${\rm E}_{7(7)}$. 
The resulting gauged field equations and Bianchi identities are globally ${\rm E}_{7(7)}$-invariant provided the constant tensor $\theta_M{}^\alpha$ is transformed under ${\rm E}_{7(7)}$ as well. 

At this point we wish to employ the same \emph{bottom-up} approach followed in the previous sections and try to interpret the most general gauged $D=4,\,{\cal N}=8$ supergravity as descending from an M-theory compactification on some generalized geometry manifold $\mathcal{M}$. 
Following \cite{Hull:2007zu} it is natural to characterize this generalized $\mathcal{M}_{56}$ as a $56$-dimensional space, dubbed the ``megatorus'', whose tangent bundle has structure group ${\rm E}_{7(7)}$. 
We assume we are compactifying on a ${\cal M}_7$, whose tangent and cotangent spaces are well defined. 
From (\ref{branch56}) it follows that the tangent space of $\mathcal{M}_{56}$ should have the form \cite{Hull:2007zu} 
\begin{equation}
	T \oplus \Lambda^2 T^* \oplus \Lambda^6 T^* \oplus \Lambda^5 T\,, \label{coordinates} 
\end{equation}
$T$, $T^*$ denoting the tangent and cotangent space of ${\cal M}_7$. 
The vielbein basis for $\mathcal{M}_{56}$ has the form: $\mathbb{E}^M=\{E^i,\,E_{ij},\,\tilde{E}_i,\,\tilde{E}^{ij}\}$. 
Similarly we define local coordinates on $\mathcal{M}_{56}$: $\mathbb{Y}^M=\{y^i,\,\mathcal{Y}_{ij},\,\widetilde{y}_i,\,\widetilde{\mathcal{Y}}^{ij}\}$, where $y^i$ are local coordinates on $\mathcal{M}_7$. 
The scalar and vector fields in the low energy $D=4,\,{\cal N}=8$ theory should then arise from the following generalized reduction ansatz 
\begin{eqnarray}
	d\mathcal{S}^2&=&\mathcal{H}_{MN}(x)\,(\mathbb{E}^M+A^M)(\mathbb{E}^N+A^N)\,,\label{genmet} 
\end{eqnarray}
where $\mathcal{H}=U^TU$ is a symmetric symplectic matrix built out of the vielbein $U$ of the scalar manifold ${\rm E}_{7(7)}/{\rm SU}_{8}$, which depends on the 70 scalars of the four dimensional theory. 
Now we wish to define a consistent deformation of the cohomology of $\mathcal{M}_{56}$ which accommodates the tensor $X_{MN}{}^P$ and which reproduces the corresponding gauged supergravity in four dimensions. 
We could naively try to write a Maurer-Cartan equation of the form: 
\begin{eqnarray}
	d\mathbb{E}^M&=&-\frac{1}{2}\,\mathcal{T}_{NP}{}^M\,\mathbb{E}^N\wedge\mathbb{E}^P=\frac{1}{2}\,X_{NP}{}^M\,\mathbb{E}^N\wedge\mathbb{E}^P\,.\label{nogood} 
\end{eqnarray}
From equation (\ref{2con3}) it follows that the above equation is not integrable and thus is inconsistent. 
This means that the one forms $\mathbb{E}^M$ are not enough for describing the cohomology of $\mathcal{M}_{56}$. 
In order to write a consistent generalization of the Maurer-Cartan equations involving the most general embedding tensor, we can introduce in the cohomology of $\mathcal{M}_{56}$ a basis of $133$ 2-forms $b_\alpha$ transforming in the adjoint representation of ${\rm E}_{7(7)}$ and write 
\begin{eqnarray}
	d\mathbb{E}^M&=&\frac{1}{2}\,X_{NP}{}^M\,\mathbb{E}^N\wedge\mathbb{E}^P-\frac{1}{2}\,\theta^{M\,\alpha}\,b_\alpha\,,\\
	\theta^{M\,\alpha}\,db_\alpha &=&-\frac{1}{2}\,\theta^{M\,\alpha}\,d_{\alpha\,NP}\,\mathbb{E}^N\wedge\left(\frac{1}{3}\,X_{RS}{}^P\,\mathbb{E}^R\wedge\mathbb{E}^S+\theta^{P\,\beta}\,b_\beta\right)\,.\label{eb} 
\end{eqnarray}
The above system of equations is indeed integrable and manifestly ${\rm E}_{7(7)}$-covariant. 

Equation (\ref{2con2}) guarantees that a symplectic rotation always exists, which can rotate the magnetic components $\theta^{\Lambda\alpha}$ of $\theta_M{}^\alpha$ to zero. 
In this electric frame the 2-forms disappear in the derivative of $\mathbb{E}^\Lambda$ which reads 
\begin{eqnarray}
	d\mathbb{E}^\Lambda&=&\frac{1}{2}\,X_{\Sigma \Gamma}{}^\Lambda\,\mathbb{E}^\Sigma\wedge\mathbb{E}^\Gamma\,,\label{neb} 
\end{eqnarray}
where we have used the property $X_\Sigma{}^{\Gamma\Lambda}=0$, following from the condition $X_{(MNP)}=0$. 
The 2-forms therefore enter only into the expression of $d\mathbb{E}_\Lambda$. 
We postpone to future work a more detailed analysis of the geometry of $\mathcal{M}_{56}$ and of the local embedding $\mathcal{M}_7\hookrightarrow \mathcal{M}_{56}$. 
The scalar potential induced by the generalized fluxes $X_{MN}{}^K$ takes a form similar to (2.12):
\begin{equation}
	V = X_{MN}{}^{R}\,X_{PQ}{}^{S}\,{\cal H}^{MP}{\cal H}^{NQ}{\cal H}_{RS}	+ 7\,X_{MN}{}^{Q}\,X_{PQ}{}^{N}\,{\cal H}^{MP} ,
    \label{sclarpotMth}
\end{equation}
where ${\cal H}_{MN}$ is now the symmetric $56\times56$ matrix that contains the 70 scalar fields of the $N=8$ theory.

The known form and geometric fluxes can be identified with different components of the most general embedding tensor under the branching of the ${\bf 912}$ with respect to ${\rm GL}(7,\mathbb{R})$: 
\begin{eqnarray}
	{\bf 912}&\rightarrow &{\bf 1}_{-7}+{\bf 1}_{+7}+{\bf 35}_{-5}+ {{\bf 35}}^\prime_{+5}+ ( {{\bf 140}}^\prime+ {{\bf 7}}^\prime)_{-3}+ ({\bf 140}+{\bf 7})_{+3}+{\bf 21}_{-1}+ {{\bf 21}}^\prime_{+1}+\nonumber\\&&{\bf 28}_{-1}+ {{\bf 28}}^\prime_{+1}+{\bf 224}_{-1}+ {{\bf 224}}^\prime_{+1}\label{branch912}\,. 
\end{eqnarray}
Each representation in the above branching is arranged in the table below together with the corresponding tensor representation. 
\begin{center}
	\begin{tabular}{||c|c|@{\hspace{1mm}}|c|c|@{\hspace{1mm}}|c|c||}\hline \rule[-3mm]{0pt}{0pt} \rule{0pt}{5mm} $\bf{1}_{+7}$ & $g_7$ & $(\bf{140} + \bf{7})_{+3}$ & $\tau^i_{jk}+\delta^i_j\tau_{k}$ &$\bf{28}_{-1}$ & $\theta_{(ij)}$ \\\hline \rule[-3mm]{0pt}{0pt} \rule{0pt}{5mm} $\bf{1}_{-7}$ & $\tilde g_7$ & $( {\bf{140}^\prime} + {\bf{7}^\prime})_{-3}$ & $Q_i^{jk}+\delta_i^j Q^k$ &$ {\bf{28}^\prime}_{+1}$ & $\xi^{(ij)}$ \\\hline \rule[-3mm]{0pt}{0pt} \rule{0pt}{5mm} $\bf{35}_{-5}$ & $h^{ijkl}$ &${\bf{224}}_{-1}$ & $f^{i}_{jkl}$ & $\bf{21}_{-1}$ & $\theta_{[ij]}$ \\\hline \rule[-3mm]{0pt}{0pt} \rule{0pt}{5mm} $ {\bf{35}^\prime}_{+5}$ & $g_{ijkl}$ &$ {\bf{224}^\prime}_{+1}$ & $R_{i}^{jkl}$ & $ {\bf{21}^\prime}_{+1}$ & $\xi^{[ij]}$\\\hline 
	\end{tabular}
	
	\smallskip Table I: flux representations under the $GL(7,{\mathbb R})$ decomposition of $E_{7(7)}$. 
\end{center}
The component $g_7$ if the flux of the 7-form field strength across $\mathbb{T}^7$, while $\tilde{g}_7$ represents the four dimensional space-time components of the 4-form field strength. 
The internal flux of the 4-form field strength is $g_{ijkl}$ while $\tau_{ij}{}^k$ is the twist of the torus. 
All the components in Table I are part of a single irreducible representation of ${\rm E}_{7(7)}$ and therefore are mapped into one another by string/M-theory dualities. 
We can also see that many new ``non-geometric'' fluxes may appear. 
\par Let us end this section by illustrating how the above scheme naturally includes the known ${\rm SO}(8)$-gauging \cite{deWit:1981eq} arising from the M-theory compactification on a seven-sphere $S^7$ and the ${\rm CSO}(p,q,r)$-gaugings ($p+q+r=8$) conjectured to originate from the non-compactification on a hyperboloid \cite{Hull:1984qz,Gibbons:2001wy}. 
We notice \cite{D'Auria:2005rv} that the tensors $\theta_{(ij)}$ ($\bf{28}_{-1}$), $\tau_{k}$ ($\bf{7}_{+3}$) and $g_7$ ( $\bf{1}_{+7}$) in Table I can be viewed as components of a symmetric $8\times 8$ matrix $\theta_{AB}=\theta_{BA}$, $A,B=1,\dots, 8$, in the ${\bf 36}$ of ${\rm SL}(8,\mathbb{R})\subset {\rm E}_{7(7)}$, according to the branching 
\begin{eqnarray}
	{\bf 36}&\rightarrow &\bf{1}_{+7}+\bf{7}_{+3}+\bf{28}_{-1}\,. 
\end{eqnarray}
We can rotate the magnetic components of the embedding tensor to zero through the symplectic rotation which derives from the dualization $A_{ij\,\mu}\leftrightarrow A^{ij}_\mu$. 
In the resulting electric frame the gauge generators have the form $X_{AB}=\{X_i,\,X_{ij}\}$ and are gauged by the electric vector potentials $A^{AB}_\mu$ in the ${\bf 28}^\prime$ of ${\rm SL}(8,\mathbb{R})$. 
Using eq. 
(\ref{neb}) we find that the electric components $\mathbb{E}^{AB}$ of the vielbein $\mathbb{E}^M$ close the following Maurer-Cartan equation 
\begin{eqnarray}
	d\mathbb{E}^{AB}&=&\theta_{CD}\,\mathbb{E}^{AC}\wedge \mathbb{E}^{DB}\,. 
\end{eqnarray}
These are the Maurer-Cartan equations of the ${\rm CSO}(p,q,r)$ group, where $p,q,r$ define the ${\rm SL}(8,\mathbb{R})$ conjugacy classes of $\theta_{AB}$. 
Indeed the matrix $\theta_{AB}$, through an ${\rm SL}(8,\mathbb{R})$ rotation, can be brought to the following diagonal form: 
\begin{eqnarray}
	\theta_{AB}&=&{\rm diag}(\stackrel{p}{\overbrace{+1,\dots , +1}},\,\stackrel{q}{\overbrace{-1,\dots , -1}},\,\stackrel{r}{\overbrace{0,\dots ,0}})\,. 
\end{eqnarray}
The case $q=r=0,\,p=8$ corresponds to the ${\rm SO}(8)$ gauging in which $\theta_{AB}=\delta_{AB}$. 

\medskip 
\section*{Acknowledgments.}

\noindent We are glad to thank R.~D'Auria, A.~Dabholkar, J.P.~Derendinger, P.M.~Petropoulos and F.~Zwirner for discussions. 
The research of G.~D.~and M.~T.~is supported by the European Union under the contract MRTN-CT-2004-005104, ``Constituents, Fundamental Forces and Symmetries of the Universe'' in which G.~D.~is associated with Padova University and M.~T.~to Torino Politecnico. 
The research of H.~S.~is supported by the Agence Nationale de la Recherche.


\end{document}